\documentclass[11pt]{article}
\usepackage{jheppub}

\usepackage{amsmath,graphicx,color,amssymb,bbm,subfigure,hyperref,array,slashed,braket,bbm,cancel}
\usepackage[all]{xy}

\def\be{\begin{equation}}
\def\ee{\end{equation}}
\def\ba{\begin{eqnarray}}
\def\ea{\end{eqnarray}}

\newcommand{\beq}{\begin{equation}}
\newcommand{\eeq}{\end{equation}}
\newcommand{\bea}{\begin{eqnarray}}
\newcommand{\eea}{\end{eqnarray}}

\newcommand{\bal}{\begin{aligned}}
\newcommand{\eal}{\end{aligned}}

\newcommand{\cN}{{\cal N}}
\newcommand{\cO}{{\cal O}}

\newcommand{\cF}{{\cal F}}
\newcommand{\cC}{{\cal C}}

\def\RR{{\rm I\kern-1.6pt {\rm R}}}

\def\ZZ{{\rm Z}\kern-3.8pt {\rm Z} \kern2pt}

\def\k{\kappa}
\def\r{\rho}
\def\a{\alpha}

\def\g{\gamma}

\def\D{\Delta}
\def\e{\epsilon}

\def\th{\theta}

\def\Om{\Omega}
\def\l{\lambda}
\def\L{\Lambda}
\def\s{\sigma}

\def\cF{{\cal F}}

\def\cN{{\cal N}}
\def\cO{{\cal O}}

\def\IR{\relax{\rm I\kern-.18em R}}

\def\inv{^{\raise.0ex\hbox{${\scriptscriptstyle -}$}\kern-.05em 1}}

\title{The $AdS_5$ non-Abelian T-dual of Klebanov-Witten as a $\mathcal{N} = 1$ linear quiver from M5-branes}

\author{Georgios Itsios$^{1}$,}
\author{Yolanda Lozano$^{2}$,}
\author{Jes\'us Montero$^{2}$,}
\author{Carlos N\'u\~nez$^{3}$}

\affiliation{$^1$ Instituto de F\'isica Te\'orica, UNESP-Universidade Estadual Paulista,
R. Dr. Bento T. Ferraz 271, Bl. II, Sao Paulo 01140-070, SP, Brazil}
\affiliation{$^2$ Department of Physics, University of Oviedo, Avda. Calvo Sotelo 18, 33007 Oviedo, Spain}
\affiliation{$^3$ Department of Physics, Swansea University, Swansea SA2 8PP, United Kingdom}

\emailAdd{gitsios@gmail.com} 
\emailAdd{ylozano@uniovi.es} 
\emailAdd{jesus.montero.a@gmail.com} 
\emailAdd{c.nunez@swansea.ac.uk}

\abstract{In this paper we study an $AdS_5$ solution constructed using non-Abelian T-duality, acting on the Klebanov-Witten background. We show that this is dual to a linear quiver with two tails of gauge groups of increasing rank. The field theory dynamics arises from a D4-NS5-NS5' brane set-up, generalizing the constructions discussed by Bah and Bobev. These realize  $\mathcal{N}=1$ quiver gauge theories built out of $\mathcal{N}=1$ and $\mathcal{N}=2$ vector multiplets flowing to interacting fixed points in the infrared.
We compute the central charge using $a$-maximization, and show its precise agreement with the holographic calculation. Our result exhibits $n^3$ scaling with the number of five-branes. This suggests an eleven-dimensional interpretation in terms of M5-branes, a generic feature of various $AdS$ backgrounds obtained via non-Abelian T-duality. 
\\[10pt]
\today
 } 

\subheader{}

\begin{document}
\def\Tr{{\textrm{Tr}}}

\maketitle


\section{Introduction}

Non-Abelian T-duality \cite{delaOssa:1992vci}, the generalization of the Abelian T-duality symmetry of String Theory to non-Abelian isometry groups, is a transformation between world-sheet field theories  known since the nineties. Its extension to all orders in $g_s$ and $\alpha^\prime$ remains however a technically-hard open problem \cite{Alvarez:1993qi}. As a result, non-Abelian T-duality does not stand as a String Theory duality symmetry, as its  Abelian counterpart does.

In the paper \cite{Sfetsos:2010uq}, Sfetsos and Thompson reignited the interest in this transformation by highlighting its potential powerful applications as a solution generating technique in supergravity. An interesting synergy between Holography (the Maldacena conjecture) \cite{Maldacena:1997re} and non-Abelian T-duality was also pointed out.  This connection was further exploited in \cite{Lozano:2012au} - \cite{variosa5}. These works have widely applied non-Abelian T-duality to generate new $AdS$ backgrounds of relevance in different contexts. While some of the new solutions avoid previously existing classifications \cite{Lozano:2012au,Macpherson:2014eza,Bea:2015fja,Lozano:2015bra}, which has led to  generalizations of existing families \cite{Apruzzi:2014qva,Bah:2015nva,Kelekci:2016uqv,Couzens:2016iot}, some others provide  the only known explicit solutions belonging to a given family \cite{Lozano:2015bra,Lozano:2015cra}, which can be used to test certain conjectures, such as 3d-3d duality \cite{3d3d}. Some of these works also put forward some ideas to define the associated holographic duals. Nevertheless, these initial attempts always encountered some technical or conceptual puzzle, rendering these proposals only partially satisfactory.

It was in the papers  \cite{Lozano:2016kum,Lozano:2016wrs,Lozano:2017ole}, where the field theoretical  interpretation of non-Abelian T-duality (in the context of Holography) was first addressed in detail. One  outcome of these works is that non-Abelian T-duality changes the dual field theory. In other words, that new $AdS$ backgrounds generated through non-Abelian T-duality  have dual CFTs different from those dual to the original backgrounds. This is possible because, contrary to its Abelian counterpart, non-Abelian T-duality has not been proven to be a String Theory symmetry.

The results in \cite{Lozano:2016kum,Lozano:2016wrs,Lozano:2017ole} open up an exciting new way to generate new quantum field theories in the context of Holography. In these examples the dual CFT arises in the low energy limit of a given Dp-NS5 brane intersection. This points to an interesting relation between $AdS$ non-Abelian T-duals and M5-branes, that is confirmed by the $n^3$ scaling of the central charges.

Reversing the logic, the understanding of the field theoretical realization of non-Abelian T-duality brings in a surprising new way (using Holography!) to extract global information about the new backgrounds. Indeed, as discussed in the various papers  \cite{Alvarez:1993qi}, one of the long-standing open problems of non-Abelian T-duality is that it fails  in determining global aspects of the dual background.

The idea proposed in \cite{Lozano:2016kum} and further elaborated in \cite{Lozano:2016wrs,Lozano:2017ole}, relies on using the dual field theory to globally define (or {\it complete}) the background obtained by non-Abelian T-duality. In this way the Sfetsos-Thompson solution \cite{Sfetsos:2010uq}, constructed acting with non-Abelian T-duality on the $AdS_5\times S^5$ background,  was completed and  understood as a superposition of Maldacena-N\'u\~nez solutions \cite{Maldacena:2000mw}, dual to a four dimensional CFT. This provides a global definition of the background and also smoothes out its singularity. This idea was also put to work explicitly in \cite{Lozano:2016wrs}  in the context of $\mathcal{N}=4$ $AdS_4$ solutions. 
In this case the non-Abelian T-dual solution was shown to arise as a patch of a geometry discussed in \cite{Assel:2011xz}, dual to the renormalization fixed point of a $T^{\hat \rho}_\rho (SU(N))$ quiver field theory, belonging to the general class introduced by Gaiotto and Witten in  \cite{Gaiotto:2008ak}. 
  
In the two examples discussed in \cite{Lozano:2016kum,Lozano:2016wrs} the non-Abelian T-dual solution arose as the result of zooming-in on a particular region of a completed and well-defined background. Remarkably, this process of  zooming-in has recently been identified more precisely as a Penrose limit of a well-known solution. The particular example studied in the paper  \cite{Lozano:2017ole}, a background with isometries $\mathbb{R}\times SO(3)\times SO(6)$,  was shown to be the Penrose limit of a given  Superstar solution \cite{Leblond:2001gn}. This provides an explicit realization of the ideas in \cite{Lozano:2016kum} that is clearly applicable in more generality. 
      
In this paper we follow the methods in \cite{Lozano:2016kum} to propose a CFT interpretation for the $\mathcal{N}=1$ $AdS_5$ background obtained in \cite{Itsios:2013wd},\cite{Macpherson:2014eza},  by acting with non-Abelian T-duality on a subspace of the Klebanov-Witten solution \cite{Klebanov:1998hh}. We show that, similarly to the examples in \cite{Lozano:2016kum,Lozano:2016wrs}, the dual CFT is given by a linear quiver with gauge groups of increasing rank. The dynamics of this quiver is shown to emerge from a D4-NS5-NS5' brane construction that generalizes the Type IIA brane set-ups  discussed by Bah and Bobev in \cite{Bah:2013aha}, 
realizing $\mathcal{N}=1$ linear quivers built out of $\mathcal{N}=1$ and $\mathcal{N}=2$ vector multiplets that flow to interacting fixed points in the infrared. These quivers can be thought of as $\mathcal{N}=1$ twisted compactifications of the six-dimensional $(2,0)$ theory on a punctured sphere, thus providing a generalization to $\mathcal{N}=1$ of the $\mathcal{N}=2$ CFTs discussed in \cite{Witten:1997sc}.

The results in this paper suggest that the non-Abelian T-dual solution under consideration could provide the first explicit gravity dual to an ordinary $\mathcal{N}=1$ linear quiver associated to a D4-NS5 brane intersection
 \cite{Bah:2013aha}. In this construction, the $\mathcal{N}=2$  SUSY D4-NS5 brane set-up associated to the Sfetsos-Thompson solution (see \cite{Lozano:2016kum}) is reduced to $\mathcal{N}=1$ SUSY through the addition of extra orthogonal NS5-branes, as in  \cite{Bah:2013aha}.
The quiver that we propose does not involve the $T_N$ theories introduced by Gaiotto \cite{Gaiotto:2009we},  and is in contrast with the classes of $\mathcal{N}=1$ CFTs constructed in \cite{Benini:2009mz,Bah:2011je,Bah:2011vv,Bah:2012dg}.
We support our proposal with the computation of the central charge associated to the quiver, which is shown to match exactly the holographic result. We also clarify a puzzle posed in \cite{Itsios:2013wd}, where the non-Abelian T-dual background was treated as a solution in the general class constructed in \cite{Bah:2011vv,Bah:2012dg}, involving the $T_N$ theories, whose corresponding central charge was however in disagreement with the holographic result. 

Before describing  the plan of this paper, let us put the present work in a wider framework, discussing  in some more detail  the general ideas behind it. 
\subsection{General framework and organization of this paper}
In the papers \cite{Itsios:2013wd}, the non-Abelian T-dual of the Klebanov-Witten background was constructed. There, it was loosely suggested that 
the dual field theory could have some relation to the ${\cal N}=1$ version of Gaiotto's CFTs. Indeed, following the ideas in \cite{Benini:2009mz}, the non-Abelian T-dual
of the Klebanov-Witten solution could be thought of as a mass deformation of the non-Abelian T-dual of $AdS_5\times S^5/\mathbb{Z}_2$, as indicated in the following diagram,
\begin{equation*}
 \xymatrixrowsep{4pc}
 \xymatrix{ 
         AdS_5 \times S^5/\mathbb{Z}_2  \ar[d]^{\textrm{mass}}  \ar[rr] && \textrm{ NATD of } AdS_5 \times S^5/\mathbb{Z}_2\ar[d]^{\textrm{mass}} \\
       AdS_5 \times T^{1,1} \ar[rr] &&  \textrm{NATD of } AdS_5 \times T^{1,1}.
           }
\end{equation*}
Nevertheless, there were many unknowns and not-understood subtle issues when the papers \cite{Itsios:2013wd} were written. To begin with, the dual CFT to the non-Abelian T-dual of $AdS_5\times S^5$ was not known, the holographic central charge of such background was not expressed in a way facilitating the comparison with the CFT result, the important role played by large gauge transformations \cite{Lozano:2013oma,Lozano:2014ata} had not been identified, etc. In hindsight, the papers \cite{Itsios:2013wd}  did open  an interesting line of research, but left various uncertainties and loose ends.

This line of investigations evolved to culminate in the works  \cite{Lozano:2016kum,Lozano:2016wrs, Lozano:2017ole}, that  gave  a precise dual field theoretical description of different backgrounds obtained by non-Abelian T-duality. This led to a field-theory-inspired {\it completion}  or {\it regularization} of the non-Abelian T-dual backgrounds. Different checks of this proposal have been performed. Most notably, the central charge is a quantity that nicely matches the field theory 
calculation with the holographic computation in the completed  (regulated) background.

In this paper we will apply the ideas of  \cite{Lozano:2016kum,Lozano:2016wrs, Lozano:2017ole} and the field theory methods of  \cite{Bah:2013aha} to the non-Abelian T-dual of the Klebanov-Witten background.  A summary of our results is:
\begin{itemize}
\item{We  perform a study of the background and its quantized charges, and  deduce the  Hanany-Witten \cite{Hanany:1996ie} brane set-up, in terms of D4 branes and two types of five-branes  NS5 and NS5'.}
\item{We  calculate the holographic central charge. This requires a regularization of the background, particularly in one of its coordinates. The regularization we adopt here is a hard-cutoff. Whilst geometrically unsatisfactory, previous experience in \cite{Lozano:2016kum} shows that this leads to sensible results, easy to compare with a field theoretical calculation.}
\item{Based on the brane set-up, we  propose a precise linear quiver field theory. This, we conjecture, is dual to the {\it regulated} non-Abelian T-dual background. We check that the quiver is at a strongly coupled fixed point by calculating the beta functions and R-symmetry anomalies. }
\item{The quiver that we propose is a generalization of those studied in  \cite{Bah:2013aha}. It can be thought of as a mass deformation of the $\mathcal{N} = 2$ quiver dual to the non-Abelian T-dual of $AdS_5 \times S^5/\mathbb{Z}_2$, that is constructed following  the ideas in \cite{Lozano:2016kum}. It is the presence of a flavor group in the CFT that {\it regulates} the space generated by non-Abelian T-duality.}
\item{We  calculate the field theoretical central charge applying the methods in  \cite{Bah:2013aha}. We find precise agreement with the central charge computed holographically for the regulated non-Abelian T-dual solution.}
\end{itemize}

In more detail, the present paper is organized as follows. In Section \ref{NATDKW}, we summarize the main properties of the solution constructed in \cite{Itsios:2013wd}. We perform a detailed study of the quantized charges, with special attention to the role played by large gauge transformations. Our  analysis suggests a D4, NS5, NS5' brane set-up associated to the solution, similar to that associated to the Abelian T-dual of Klebanov-Witten, studied in \cite{Dasgupta:1998su,Uranga:1998vf}. In Section \ref{subsec_linear_quivers} we summarize the brane set-up and $\mathcal{N}=1$ linear quivers of \cite{Bah:2013aha}, which we use in Section \ref{quiverNATD} for the proposal of a linear quiver that, we conjecture, is dual to the {\it regulated} version of the non-Abelian T-dual solution  of  $AdS_5\times T^{1,1}$. We provide support for our proposal with the detailed computation of the (field theoretical) central charge which we show to be in full agreement with the (regulated) holographic result. We give an interpretation for the field theory dual to our background in terms of a mass deformation of the $\mathcal{N}=2$ CFT  associated to the non-Abelian T-dual of $AdS_5\times S^5/\mathbb{Z}_2$. This suggests the geometrically sensible way of completing our background. 
Section \ref{section5} contains a discussion where we further elaborate on the relation between our proposal and previous results in \cite{Itsios:2013wd}. We also resolve a puzzle raised there regarding the relation between the non-Abelian T-dual solution and the solutions in \cite{Bah:2012dg}.
Concluding remarks and future research directions are presented in Section \ref{conclusions}.
Detailed appendices complement our presentation. In Appendix \ref{appendixgmsw}, we explicitly calculate the differential forms showing that the non-Abelian T-dual solution 
fits  in the classification of \cite{Gauntlett:2004zh}, for ${\cal N}=1$ SUSY spaces with an $AdS_5$-factor. 
Appendix \ref{appendixa} studies  in detail the relation between the non-Abelian T-dual solution and its (Abelian) T-dual counterpart. Finally in Appendix \ref{appendixb} we present some field theory results relevant for the analysis in Section \ref{quiverNATD}.






\section{The non-Abelian T-dual of the Klebanov-Witten solution} \label{NATDKW}

In this section we summarize the Type IIA supergravity solution obtained after a non-Abelian T-duality transformation acts  on the $T^{1,1}$ of the Klebanov-Witten background \cite{Klebanov:1998hh}. This solution was first derived in \cite{Itsios:2013wd}.
It was later studied in \cite{Macpherson:2014eza} where a more suitable set of coordinates was used.
We start by introducing our conventions for the background and by summarizing the calculation of the holographic central charge of the $AdS_5\times T^{1,1}$ solution.

\subsection{The $AdS_5\times T^{1,1}$ solution}

The metric is given by,
\begin{align}\label{eq:metric_KW}
ds^2&=ds^2_{AdS_5}+L^2 \, ds_{T^{1,1}}^2,
\\[5pt]
ds^2_{AdS_5} & =\frac{r^2}{L^2} \, dx^2_{1,3} + \frac{L^2}{r^2} \, dr^2, \qquad
ds_{T^{1,1}}^2=\lambda_1^2(\sigma_{\hat{1}}^2+
\sigma_{\hat{2}}^2)+\lambda_2^2(\sigma_{1}^2+\sigma_{2}^2)+
\lambda^2(\sigma_3+\cos\theta_1 \, d\phi_1)^2, \nonumber
\end{align}
where $\lambda^2=\frac{1}{9},\ \lambda_1^2=\lambda_2^2=\frac{1}{6}$ and
\begin{eqnarray}
\sigma_{\hat{1}} &=& \sin\theta_1 \, d\phi_1, \qquad   \sigma_{\hat{2}}=d\theta_1,  \nonumber
 \\[5pt]
\sigma_1&=& \cos\psi \, \sin\theta_2 \, d\phi_2 - \sin\psi \, d\theta_2, \qquad \sigma_2 = \sin\psi \, \sin\theta_2 \, d\phi_2 + \cos\psi \, d\theta_2,
 \\[5pt]
\sigma_3 &=& d\psi+\cos\theta_2 \, d\phi_2. \nonumber
\end{eqnarray}
The background includes a constant dilaton and a self-dual RR five-form,
\beq
F_5=\frac{4}{g_s \, L} \, \Big[\text{Vol} \big( AdS_5 \big)- {L^5} \, \text{Vol} \big( T^{1,1} \big) \Big].
\eeq
The associated charge is given by
\beq
\frac{1}{2 \, \kappa_{10}^2 \, T_{D3}}\int_{T^{1,1}}F_5=N_{3}\,.
\eeq
Using that $2 \, \kappa_{10}^2 \, T_{Dp}= (2\pi)^{7-p} \, g_s \, \alpha'^{\frac{7-p}{2}}$ this leads to a quantization of the size of the space,
\beq
\label{LN3}
L^4= \frac{27}{4} \, \pi \, g_s^2 \, \alpha'^2 \, N_{3}\,.
\eeq

To calculate the holographic central charge of this background, we use the formalism developed in  \cite{Klebanov:2007ws,Macpherson:2014eza}.
Indeed, for a generic  background  and dilaton of the form,
\begin{equation}
 ds^2=a(r,\theta^i) \, \Big[dx_{1,d}^2 +  b(r) \, dr^2\Big] + g_{ij}(r,\theta^i) \, d\theta^i \, d\theta^j,  \;\;\; \Phi(r,\theta^i),
\label{xxy}
 \end{equation}
 we define the quantities $\hat{V}_{int}, \, \hat{H}$ as, 
 \begin{equation}
 \label{cc2}
 \hat{V}_{int}=\int\! d\theta^i\sqrt{\det [g_{ij}] \, e^{-4\Phi} \, a^{d}}\,,\qquad \hat{H}=\hat{V}_{int}^2\,.
\end{equation}
The holographic central charge for the $(d+1)$-dimensional QFT is calculated as,
\beq
c=\pi \, d^d \, \frac{b^{d/2 } \hat{H}^{\frac{2d+1}{2}}}{G_{N,10} \, \big( \hat{H}' \big)^d}\,,   \;\;\;\;\; G_{N,10}= 8 \, \pi^6 g_s^2 \, \alpha'^4 .
\label{formulacentralcharge}
\eeq
Using these expressions for the background in eq.(\ref{eq:metric_KW}), we have
\beq
\label{cc1}
a=\frac{r^2}{L^2},\;\; b=\frac{L^4}{r^4}, \;\; d=3, \;\; \sqrt{e^{-4\Phi} \, \det[g_{ij}] \, a^3}= g_s^{-2} L^2 r^3 \lambda \, \lambda_1^4 \, \sin\theta_1 \, \sin\theta_2 \, .
\eeq
After some algebra, we obtain the well-known result  \cite{Gubser:1998vd},
\beq\label{eq:ccKW}
c_{KW}={\pi}{\frac{L^8}{108\pi^3 g_s^4\alpha'^4}= \frac{27}{64} N_{3}^2}  \,.
\eeq

We now study the action of non-Abelian T-duality on one of the $SU(2)$ isometries displayed by the background in eq.\! \eqref{eq:metric_KW}. We use the notation and conventions in
 \cite{Macpherson:2014eza}.
\subsection{The non-Abelian T-dual solution}
The NS-NS sector of the non-Abelian T-dual solution constructed in \cite{Itsios:2013wd,Macpherson:2014eza} is composed of a metric, a NS-NS two-form and a dilaton. Using the variables in  \cite{Macpherson:2014eza}, the metric reads\footnote{Henceforth we use the rescaling $\rho \longrightarrow \frac{L^2}{\a'}\,\rho$ so that all factors in the internal metric scale with $L^2$. We also substitute $\l_2=\l_1$ for convenience.}, 
\begin{equation}\label{eq:metric_NATD}
 \begin{aligned}
  d \hat{s}^2 & = \frac{r^2}{L^2} dx^2_{1,3} + \frac{L^2}{r^2} dr^2 + L^2 \l_1^2 \big(   d\theta_1^2 + \sin^2\theta_1 d\phi^2_1  \big) + \frac{L^2}{Q} \Big[  \lambda_1^4 \ \big( \cos\chi \ d\rho - \rho \sin\chi \ d\chi   \big)^2
  \\[5pt]
  & +  \lambda^2 \lambda_1^2  \big(   \sin\chi \ d\rho + \rho \cos\chi \ d \chi  \big)^2 +  \lambda^2 \lambda_1^2 \,\rho^2 \sin^2 \chi \  \big( d\xi + \cos\theta_1 d\phi_1 \big)^2 + \rho^2 d\rho^2
  \Big] \ .
 \end{aligned}
\end{equation}
The NS two-form is,
\begin{equation}\label{eq:B2_NATD}
 \begin{aligned}
   B_2 & = \frac{L^2\rho^2 \sin\chi}{2 Q} \Big[ \big( \lambda^2 - \lambda_1^2 \big) \sin 2\chi \ d\xi \wedge d\rho + 2 P \, \rho \ d\xi \wedge d\chi  \Big]
\\[5pt]
         & - \frac{L^2\lambda^2 \cos\theta_1}{Q} \Big[   \big(   \lambda_1^4 + \rho^2   \big) \cos\chi d\rho \wedge d\phi_1 -  \lambda_1^4 \ \rho \sin\chi \ d\chi \wedge d\phi_1   \Big] \ ,
 \end{aligned}
\end{equation}
and the dilaton is given by,
\begin{equation}\label{eq:NATD_dilaton}
 e^{-2 \hat\Phi} = \frac{L^6 \ Q}{g_s^2 \ \alpha'^3} \ .
\end{equation}
For convenience we have defined the following functions,
\begin{equation}
 Q = \lambda^2 \lambda_1^4 +  \rho^2 P \ , \qquad P = \lambda^2 \cos^2\chi  + \lambda_1^2 \sin^2\chi = \lambda_1^2 + \big(  \lambda^2 - \lambda_1^2   \big) \cos^2\chi \ .
\end{equation}
This solution is supported by a set of RR fluxes which read,
\begin{equation}\label{eq:RR_NATD}
 \begin{aligned}
  F_2 & = - \frac{4 \, L^4 \lambda \, \lambda_1^4}{g_s \, \alpha'^{3/2}} \, \sin\theta_1 \, d\theta_1 \wedge d\phi_1 \ ,
  \\[5pt]
  F_4 & = - \frac{2 \, L^6 \lambda \, \lambda_1^4}{g_s\, \alpha'^{3/2}  Q} \, \rho^2 \sin\chi \, \sin\theta_1 \, d\theta_1 \wedge d\phi_1\wedge \Big[   \big(  \lambda^2 - \lambda_1^2  \big)  \sin 2\chi \, d\xi \wedge d\rho + 2 \, P \, \rho \, d\xi \wedge d\chi   \Big]
  \\[5pt]
  & = B_2 \wedge F_2 \ .
 \end{aligned}
\end{equation}
The higher rank RR fields which are related to the previous ones through $F_p = \big(   - 1  \big)^{[p/2]} \star F_{10-p}$ read,
\begin{equation}
 \begin{aligned}
  F_6 & = - \frac{4 \, L^3}{g_s \, \alpha'^{3/2}} \, \rho \, \textrm{Vol}_{AdS_5} \wedge d\rho \ ,
  \\[5pt]
  F_8 & = - \frac{4 \, L^5 \lambda^2 \lambda_1^4}{g_s \,\alpha'^{3/2} \, Q} \, \rho^2 \, \sin\chi \, \textrm{Vol}_{AdS_{5}} 
  \wedge d\rho \wedge d\chi \wedge \big(   d\xi + \cos\theta_1 \, d\phi_1  \big) \ . 
 \end{aligned}
  \label{lalal}
\end{equation}
The associated RR potentials $C_1$ and $C_3$, defined through the formulas
 $F_2 = dC_1$  and $ F_4 = dC_3 - H_3 \wedge C_1 $,
%
are given by,
\begin{equation}
 \label{RRpotentials}
 \begin{aligned}
  C_1 & = \frac{4 \, L^4 \, \l \, \l^4_1}{g_s\,\a'^{3/2}} \, \cos\th_1 \, d\phi_1 \ , 
  \\[10pt]
  C_3 & = \frac{2 \, L^6 \, \l \, \l^4_1}{g_s\,\a'^{3/2}  Q} \, \r^2 \, \cos\th_1 \, \sin\chi \, \Big[   \big(  \l^2_1 - \l^2  \big) \, \sin 2\chi \, d\r \wedge d\xi - 2 \, P \, \r \, d\chi \wedge d\xi  \Big] \wedge d\phi_1
  \\[10pt]
          & = B_2 \wedge C_1 \ .
 \end{aligned}  
\end{equation}
In the papers \cite{Itsios:2013wd} this solution of the Type IIA equations of motion was shown to preserve $\mathcal{N} = 1$ supersymmetry. In the coordinates used in this paper the Killing vector $\partial_\xi$ is dual to the R-symmetry of the CFT.

In Appendix \ref{appendixgmsw} we promote the background in eqs.(\ref{eq:metric_NATD})-(\ref{RRpotentials}) to a solution of eleven-dimensional supergravity. We show that 
this background fits in the classification of ${\cal N}=1$ $AdS_5$ solutions in M-theory of  \cite{Gauntlett:2004zh}. We write in detail the forms satisfying a set of differential relations and  define   the $SU(2)$-structure. The eleven dimensional lift suggests that this solution is associated to M5-branes wrapped on a spherical 2d manifold. 
We discuss this picture further in Section \ref{section5}.

As indicated, one goal of this paper is to propose a conformal field theory dual to the Type IIA non-Abelian T-dual solution. We will do this by combining different insights coming from 
the large $\rho$-asymptotics, the quantized charges and the calculation of field theoretical observables using the background.

\subsubsection{Asymptotics}
In complicated systems, like those corresponding to intersections of branes, it is often illuminating to consider the asymptotic behavior of the background.
In the case at hand, for the background in eqs.(\ref{eq:metric_NATD})-(\ref{RRpotentials}), we consider the leading-order behavior of the solution, when $\rho\to\infty$.
This  allows us to {\it read}  the brane intersection that in the decoupling limit and for a very large number of branes
generates the solution.

Indeed, for $\rho\to\infty$, the leading behavior of the  NS-fields
is
\begin{eqnarray}\label{metricasympt}
 ds^2 & \approx & ds^2_{AdS_5} + L^2 \, \lambda_1^2 \, \Bigl[ d\Omega_2^2(\theta_1,\phi_1)+  \Big(  d\chi^2+ \frac{\lambda^2 \sin^2\chi}{P(\chi)} \, \big( d\xi +\cos\theta_1d\phi_1 \big)^2  \Big)
+\frac{d\rho^2}{\lambda_1^2 P(\chi)}\Bigr] \;,\nonumber
\\[5pt]
 B_2 & \approx & -L^2\rho \Bigg[ d\Omega_2(\chi,\xi)+ \frac{\lambda^2\cos\chi}{P(\chi)} \, d\Omega_2(\theta_1,\phi_1) - \lambda^2 \, \cos\theta_1 \, \partial_\chi
\Bigg(  \frac{\cos\chi}{P(\chi)} \Bigg) \, d\chi\wedge d\phi_1  \Bigg] \nonumber
\\[5pt]
& & \qquad \qquad \qquad  +\frac{L^2 \sin\chi}{2\,P(\chi)} \, \big( \l^2-\l_1^2 \big) \, \sin 2\chi \, d\xi\wedge d\rho \;,
\\[5pt]
 e^{-2\phi} & \approx & \frac{L^6}{g_s^2 \alpha'^3} \, P(\chi) \, \rho^2 \nonumber \;,
\end{eqnarray}
where we have performed a gauge transformation in $B_2$, of the form $B_2+d\Lambda_1$, with 
\[ \Lambda_1= L^2 \lambda^2 \rho \cos\theta_1 \Bigg(  \frac{\cos\chi}{P(\chi)}\Bigg)   d\phi_1 \;. \]

Intuitively, this result suggests that we have two different types of NS-five branes. One type of five-branes (which we  refer to as {\it NS}\,) 
extend along $\mathbb{R}^{1,3}\times S^2(\theta_1,\phi_1)$. The 
second type of five branes (referred to as {\it NS'}\,) extend along $\mathbb{R}^{1,3}\times \tilde{S}^2(\chi,\xi)$ . To preserve SUSY, the spaces $S^2(\theta_1,\phi_1)$ and $\tilde{S}^2(\chi,\xi)$ 
are fibered by the monopole gauge field $A_1=\cos\theta_1 d\phi_1$. This fibration is also reflected in  the $B_2$-field, that contains a term that mixes the spheres.

The asymptotics of the RR-fields can be easily read from eq.(\ref{lalal}). Indeed, the expression $F_6=dC_5$, generates asymptotically
 $C_5\approx \rho r^4 dx_{1,3}\wedge d\rho$. This suggests an array of D4 branes extended along the directions $\mathbb{R}^{1,3}\times \rho$.
 D6 branes appear due to the presence of the $B_2$-field, that blows up the D4 branes due to the Myers effect \cite{Myers:1999ps}.
 
 In summary, the asymptotic analysis suggests that the background in eqs.(\ref{eq:metric_NATD})-(\ref{RRpotentials}), is generated
 in the decoupling limit of an intersection of NS5-NS5'-D4 branes. This will be confirmed by the calculation of the  quantized charges associated to this solution.


\subsubsection{Quantized charges}\label{quantized-charges}

In the papers \cite{Lozano:2016kum,Lozano:2016wrs}, the brane set-ups encoding the dynamics of the  CFTs dual to the corresponding non-Abelian T-dual backgrounds were proposed after a careful analysis of the quantized charges.
The charges that are relevant for the study of the non-Abelian T-dual of the Klebanov-Witten background are those related to $D4$, $D6$ and $NS5$ branes.  Based on this analysis we will propose an array of branes, from which the dynamics of a linear quiver with gauge groups of increasing rank will be obtained.
%

For $D6$ branes the Page charge reads, 
\begin{equation}
 Q_{D6} = \frac{1}{2 \, \kappa^2_{10} \, T_{D6}} \int_{(\theta_1, \phi_1)} F_2 = \frac{8 \, L^4 \lambda \, \lambda_1^4}{g_s^2 \ \alpha'^2} = \frac{2}{27} \frac{L^4}{g_s^2 \, \alpha'^2} = N_6 \ ,
\end{equation}
where we have absorbed an overall minus sign by choosing an orientation for the integrals.
Imposing the quantization of the $D6$ charge, the $AdS$ radius $L$ is quantized in terms of $N_6$, 
\begin{equation}\label{cuantizationdespuesnatdkw}
 L^4 = \frac{27}{2} \, g_s^2 \, \alpha'^2 N_6 \ .
\end{equation}
This relation differs from that for the original background, see eq. \eqref{LN3}, which is a common feature already observed in the bibliography \cite{Lozano:2013oma}. 

In turn, the Page charge associated to $D4$-branes vanishes, 
\begin{equation}
 Q_{D4} = \frac{1}{2 \, \kappa^2_{10} \, T_{D4}} \int_{M_4} \big( F_4 - F_2 \wedge B_2 \big) = 0 \ .
\end{equation}
This charge becomes however important in the presence of large gauge transformations,
\begin{equation}
 B_2 \rightarrow B_2 + \Delta B_2 \ ,
\end{equation}
under which the Page charges transform as,
\begin{equation}
 \Delta Q_{D4} =  - \frac{1}{2 \, \kappa^2_{10} \, T_{D4}} \int_{M_4} F_2 \wedge \Delta B_2 \ , \qquad \Delta Q_{D6} = 0 \ .
\end{equation}

Indeed, consider a four-manifold $M_4 = [\theta_1, \phi_1] \times \Sigma_2$, with the two-cycle given by $\Sigma_2=[\chi,\xi]$\footnote{Note that this 2-cycle vanishes at $\rho\to 0$, while at $\rho\to\infty$ it is {\it almost} a two sphere of finite size.}. Under a large gauge transformation of the form,
\begin{equation}
\label{largegauge}
 \Delta B_2 = - n \, \pi \, \alpha' \sin\chi \, d\chi \wedge d\xi \ , \qquad d\big[   \Delta B_2  \big] = 0 \ ,
\end{equation}
the Page charges transform as
\begin{equation}
\label{D4large}
 \Delta Q_{D4} = n \, N_6 \ , \qquad \Delta Q_{D6} = 0 \ .
\end{equation}
The first relation shows that $n$ units of D4-brane charge are induced in each D6-brane. Conversely, $nN_6$ D4-branes can expand in the presence of the $B_2$ field given by eq. (\ref{largegauge}) into $N_6$ D6-branes wrapped on $\Sigma_2$, through Myers dielectric effect.
Consider now the (conveniently normalized) integral of the $B_2$ field, given by eq. (\ref{eq:B2_NATD}), along the non-trivial 2-cycle $\Sigma_2=[\chi,\xi]$. Following
the paper \cite{Lozano:2014ata}, this must take values in the interval $[0,1]$\footnote{A physical interpretation of this condition in terms of a fundamental string action was presented in \cite{Macpherson:2015tka}.}. Imposing this condition implies that $|b_0|\leq 1$ with,
\begin{equation}
\label{b0}
 b_0 = \frac{1}{4 \, \pi^2 \, \alpha'} \int_{\Sigma_2} B_2 =- \frac{1}{\pi} \frac{L^2}{\a'} \Bigg[   \rho - \frac{\sqrt{2}}{6\sqrt{1+54 \, \rho^2}} \tanh^{-1}\!\! \left(   \frac{3\sqrt{2}\,\rho}{\sqrt{1+54 \, \rho^2}} \right)  \Bigg] \ .
\end{equation}
The asymptotic behavior of $b_0$ for small and large values of $\rho$ is given by,
\begin{equation}
 \begin{aligned}
  & b_0 = - \frac{48 \, L^2 \rho^3}{\pi \alpha'} + \cO(\rho^5) \ , \qquad \rho \ll 1 \ ,
  \\[5pt]
  & b_0 = -\frac{L^2}{\a'}\frac{\rho}{\pi} + \cO\Big(   \frac{1}{\rho}  \Big) \ , \qquad \rho \gg 1 \ .
 \end{aligned}
\end{equation}
The expression given by eq.(\ref{b0}) is monotonically increasing for all $\rho\in [0,\infty)$, and takes the value $|b_0|=1$ only once. In order to bring the function $|b_0(\rho)|$ back to the interval $[0,1]$ we need to perform a large gauge transformation of the type defined in eq. (\ref{largegauge}), whenever $|b_0(\rho_n)| = n, \, n \in \mathbb{N}^{}$. 
The number of D4-branes in the configuration then increases by  a multiple of $N_6$, as implied by eq. (\ref{D4large}), each time we cross the position $\rho=\rho_n$.

The form of the $B_2$ potential in eq. \eqref{eq:B2_NATD} suggests that  it is also possible to take a different 2-cycle, 
\beq
\Sigma'_2=[\theta_1,\phi_1]_{\chi=0}\;,
\eeq
which is a rounded $S^2(\th_1, \phi_1)$ at $\chi=0$. As in the  case analyzed above, large gauge transformations are needed as we move in $\rho$ in order to render $b_0$ in the fundamental region, $b_0\in [0,1]$. This shift does not modify however the number of D4 or D6-branes, while it induces NS5-brane charge (we call these NS5' for later convenience) in the configuration.

Indeed, let us  discuss the NS5-brane charges associated to the solution.
Let us first consider the three-cycle,
\begin{equation}
 \Sigma_3 = [\rho,\chi,\xi] \ ,
\end{equation}
built out of the first 2-cycle $\Sigma_2=[\chi,\xi]$ and the $\rho$-coordinate.
Taking into account the expression for the $B_2$ field given by eq. (\ref{eq:B2_NATD}) one finds,
\begin{equation}
 H_3 \big|_{\Sigma_3} = L^2 \, \Bigg[    \frac{(\lambda^2 - \lambda_1^2) \, \rho^2}{2} \, \partial_{\chi} \Big(   \frac{\sin\chi \sin 2\chi}{Q}  \Big) -  P \sin\chi \ \partial_{\rho} \Big(  \frac{\rho^3}{Q}  \Big)    \Bigg] d\rho \wedge d\chi \wedge d\xi \ .
\end{equation}
The first term does not contribute to the charge, which reads,
\begin{equation}
 Q_{NS5} = \frac{1}{4 \, \pi^2 \, \alpha'} \int_{(\rho, \chi, \xi)} H_3 = - \frac{1}{4 \, \pi^2 \, \alpha'} \, 2 \, \pi \, L^2 \rho_n^3 \int\limits_{0}^{\pi} \frac{P}{Q} \, \sin\chi \, d\chi = b_0(\rho_n) = n \ .
\end{equation}
This calculation shows that we have $ n$ $NS5$ branes for $\rho\in[0, \rho_n]$. 
If, on the other hand, we take the cycle defined by 
\begin{equation}
 \Sigma_3' = [\rho,\phi_1,\theta_1]_{\chi=0} ,
\end{equation}
 we find that $\rho'_n = n \,\pi \, \alpha' /L^2$ and that a new NS5' brane is created each time we cross these values  $\rho'_n$ for $n=1,2,\dots$.

\begin{figure}[ht]
	\begin{center}
		\includegraphics[width=15cm]{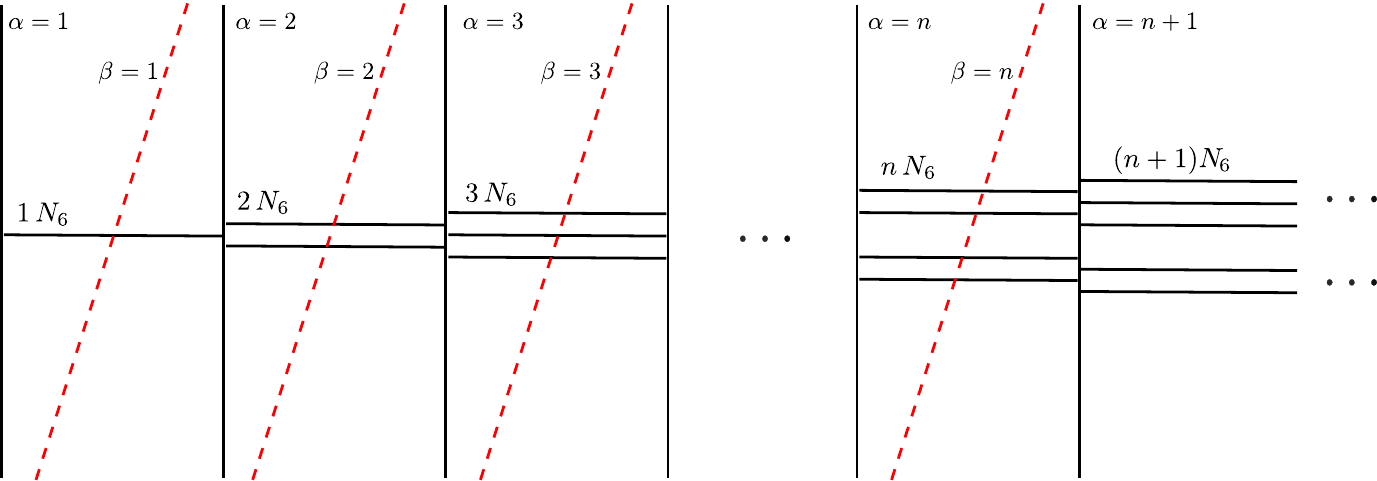}
	\end{center}
	\caption{Brane set-up consistent with the quantized charges of the non-Abelian T-dual solution, consisting on $\alpha=1,2,\ldots,n+1$ NS5-branes (vertical black lines), $\beta=1,2,\ldots,n$ NS5'-branes (tilted red dashed lines) and $m\,N_{6}$ D4-branes (horizontal lines), where $m=1,2,\ldots,n+1$ changes by one each time a NS5-brane is crossed.}
	\label{fig:NATDbranes}	
\end{figure}

The conclusion of this analysis is that one can define {\it two types} of NS5-branes in the non-Abelian T-dual background: NS5-branes located at  $\rho_n$ and transverse to ${\tilde S}^2(\chi,\xi)$, and NS5'-branes located at $\rho'_n = n \,\pi \, \alpha' /L^2$ and transverse to $S^2(\th_1,\phi_1)$. These branes
are localized in the $\rho$ direction, such that a NS5'-brane lies in between each pair of NS5-branes, as illustrated in Figure \ref{fig:NATDbranes}. Further, as implied by eq. (\ref{D4large}), $N_6$ D4-branes are created each time a NS5-brane is crossed.
This brane set-up will be the basis of our proposed quiver in Section \ref{quiverNATD},  and will be instrumental in defining the dual CFT of the non-Abelian T-dual solution. As we will see, it will allow us  to identify the global symmetries and the parameters characterizing  the associated field theory. 

Let us study now an important field theoretical quantity, calculated from the Type IIA solution, the central charge.


\subsubsection{Central charge}
\label{centralcharge}

In this section, we compute the holographic central charge associated to the non-Abelian T-dual solution in eqs.(\ref{eq:metric_NATD})-(\ref{RRpotentials}). This will be  the main observable to check the validity of the $\mathcal{N}=1$ quiver  proposed in Section \ref{quiverNATD}.

We must be careful about the following subtle point.
The calculation of the quantity $\hat{V}_{int}$ in eq.(\ref{cc2}),
will involve an integral in the $\rho$-direction of the metric in eq.(\ref{eq:metric_NATD}). The range of this coordinate is not determined by the process of non-Abelian T-duality (the global issues we referred to in the Introduction). If we take $0\leq\rho<\infty$, we face the problem that the central charge will be strictly infinite. A process of {\it regularization} or {\it completion} of the background of eqs.(\ref{eq:metric_NATD})-(\ref{RRpotentials}) is needed. In this paper we choose to end the space with a hard cut-off, namely $0\leq \rho\leq \rho_n$. We do know that this is geometrically unsatisfactory. Nevertheless, the field theoretical analysis of Section \ref{quiverNATD} will teach us that a flavor group, represented by D6 branes added to the background of eqs.(\ref{eq:metric_NATD})-(\ref{RRpotentials}), should end the space in the correct fashion. Previous experience  \cite{Lozano:2016kum} tells us that the hard-cutoff used here does capture the result for the holographic central charge that is suitable to compare with the field theoretical one found in Section \ref{quiverNATD}. 

We then proceed, by considering the metric in eq.(\ref{eq:metric_NATD}), the dilaton in eq.(\ref{eq:NATD_dilaton}) and eqs.(\ref{xxy})-(\ref{formulacentralcharge}). We obtain,
\beq\label{eq:ccNATD}
c_{KWNATD}=\frac{9 \, L^6 \, \big( \rho_{b}^3-\rho_a^3 \big)}{64 \, \pi^3 \, \alpha'^3} \, N_{6}^2,
\eeq
where we have integrated $\rho$ between two arbitrary values $[\rho_a, \rho_b]$. We have also used the quantization condition of eq.(\ref{cuantizationdespuesnatdkw}).
For $\rho \in \big[ 0, n\,\pi \, \alpha' / L^2 \big)$  this gives
\beq\label{eq:cc_NATD_0n}
c_{KWNATD}^{(0,n)}= \frac{9}{64} \, n^3 N_{6}^2  \;.
\eeq
On the other hand, for $\rho \in \big[ n \, \pi \, \alpha' /L^2, \, (n+1) \, \pi \, \alpha' / L^2 \big)$ we obtain, 
\beq
c_{KWNATD}^{(n,n+1)}= \frac{9}{64} \, N_{6}^2 \, \big( 3 \, n^2 + 3 \, n + 1 \big) .
\label{mama}
\eeq
This becomes 
$c_{KWNATD}^{(n,n+1)}=\frac{27}{64} \, N_{4}^2$ in the large $n$ limit, where $\rho_n'=\rho_n$ and we can use that $N_4=nN_6$ in the $\big[ n \, \pi \, \alpha' /L^2, \, (n+1) \, \pi \, \alpha' / L^2 \big)$ interval. Interestingly, this expression coincides with the central charge of the Abelian T-dual of the Klebanov-Witten background, that we discuss in detail in Appendix \ref{appendixa}. This  is that of the original background -- see eq.(\ref{eq:ccKW}), with $N_{3}$ replaced by $N_{4}$,
\beq
\label{abeliancc}
c_{KWATD}= \frac{27}{64} \, N_{4}^2 .
\eeq
For completeness, we also reproduce in Appendix \ref{ATDZk} this value of the central charge from the field theory, using $a$-maximization.
This matching between the central charges of non-Abelian and Abelian T-duals was found in previous examples \cite{Lozano:2016kum,Lozano:2016wrs}.

Next, we review aspects of the $\mathcal{N}=1$ quivers discussed in \cite{Bah:2013aha}. These will be the basis of the quiver proposed to describe the field theory associated to the non-Abelian T-dual solution. In Section \ref{quiverNATD}, the holographic result in eq.(\ref{eq:cc_NATD_0n}) will be found by purely field theoretical means.

\section{Basics of Bah-Bobev 4d $\cN=1$ theories}\label{subsec_linear_quivers}
In this section, we provide a summary of the results in  \cite{Bah:2013aha}, which will be instrumental for our proposal of a field theory dual to the background in eqs.(\ref{eq:metric_NATD})-(\ref{RRpotentials}).

\subsection{$\mathcal{N}=1$ linear quivers}

In \cite{Bah:2013aha}, Bah and Bobev introduced $\cN=1$ linear quiver gauge theories built out of $\mathcal{N}=2$ and $\mathcal{N}=1$ vector multiplets and ordinary matter multiplets. These theories  were argued  to flow to interacting 4d $\cN=1$ SCFTs in the infrared. 
\begin{figure}[ht]
	\begin{center}
		\includegraphics[width=15cm]{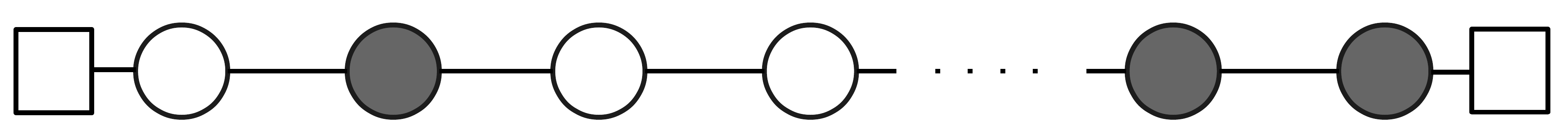}
	\end{center}
	\vspace{-0.5cm}
		\caption{General linear quiver in \cite{Bah:2013aha}. Shaded (unshaded) circles represent $SU(N)$, $\mathcal{N}=1$  ($\mathcal{N}=2$) vector multiplets. Lines between them represent bifundamentals of $SU(N)\times SU(N)$. The boxes at the two ends represent $SU(N)$ fundamentals.}
	\label{fig:BB_linear_quiver}
\end{figure}
They consist of products of $\ell-1$ copies of $SU(N)$ gauge groups, with either $\cN=1$ (shaded) or $\cN=2$ (unshaded) vector multiplets -- see Figure \ref{fig:BB_linear_quiver}.
 Let $n_1$ be the number of $\cN=1$ vector multiplets and $n_2$ the number of $\cN=2$ vector multiplets.
There are also $\ell-2$ bifundamental hypermultiplets of $SU(N)\times SU(N)$, depicted in Figure \ref{fig:BB_linear_quiver} as lines between the nodes, and two sets of $N$ hypermultiplets transforming in the fundamental of the two end $SU(N)$ gauge groups. Thus, there are in total $\ell -1=n_1+n_2$ gauge groups and $\ell$ matter multiplets. 
The total global symmetry is,
\[ SU(N)\times SU(N)\times U(1)^{\ell+n_2}\times U(1)_R , \]
corresponding to the $SU(N)$ flavor symmetries acting on the end hypermultiplets, the $U(1)$ flavor symmetry acting on each of the $\ell$ hypermultiplets, the $U(1)$ flavor acting on the chiral adjoint superfields (there are as many as $\mathcal{N}=2$ vector multiplets) and the 
R-symmetry. Out of these $U(1)'s$ only a certain non-anomalous linear combination will survive in the IR SCFT. Similarly, the fixed point R-charge is computed through a-maximization \cite{Intriligator:2003jj} as a non-anomalous linear combination of the $U(1)$'s and $U(1)_R$.

As shown in \cite{Bah:2013aha}, it is convenient to assign a charge $\sigma_i=\pm 1$ to each matter hypermultiplet, with the rule that $\mathcal{N}=1$ vector multiplets connect hypermultiplets with opposite sign, while $\mathcal{N}=2$ vector multiplets connect hypermultiplets with the same sign.
Let $p$ be the number of hypermultiplets with $\sigma_i=+1$ and $q=\ell-p$ those with $\sigma_i=-1$,  and let us introduce the twist parameter $z$,
\beq\label{eq:def_z}
z= \frac{p-q}{\ell} \;.
\eeq
Thus, $z=\pm1$ corresponds to a quiver with only $\cN=2$ nodes, involving hypermultiplets of the same charge. $z=0$ corresponds in turn to a quiver with the same number of hypermultiplets of each type, so it includes the quiver with only $\cN=1$ nodes. We will focus on $0\leq z \leq 1$ ($q\leq p$) without loss of generality. We also introduce $\k=(\sigma_0+\sigma_l)/2$, which can take values $\k=-1,0,+1$. This will later be associated to the type of punctures on the Riemann surface on which  M5-branes are wrapped.


In a superconformal fixed point the $a$ and $c$ central charges can be computed from the 't Hooft anomalies associated to the R-symmetry \cite{Anselmi:1997am},
\begin{equation}
a=\frac{3}{32} \, \big( 3 \, {\rm Tr}\,R^3-{\rm Tr}\,R \big) \, , \qquad c=\frac{1}{32} \, \big( 9 \,{\rm Tr}\,R^3 - 5 \, {\rm Tr}\,R \big)\, ,
\end{equation}
where the R-symmetry is given by
\begin{equation}
R_\epsilon=R_0+\frac{1}{2} \, \epsilon \, {\cal F}\, ,
\end{equation}
and $R_0$ is the anomaly free R-symmetry, ${\cal F}$ is the non-anomalous global $U(1)$ symmetry and $\epsilon$ is a number that is determined by $a$-maximization \cite{Intriligator:2003jj}. 
This was used in \cite{Bah:2013aha} to compute 
the $a$ and $c$ central charges associated to the general quiver represented in Figure \ref{fig:BB_linear_quiver}. Their values were shown to depend only on the 
set of parameters $\{\kappa,z,\ell,N\}$. It was then conjectured that all quivers with the same $\{\kappa,z,\ell,N\}$ should be dual to each other and flow to the same SCFT in the infrared. 
Moreover, for $\ell\rightarrow\infty$ the two central charges were shown to agree. Therefore, in this limit the quivers can admit holographic $AdS$ duals. In Section \ref{quiverNATD} we will provide  a variation of these $\mathcal{N}=1$ quivers for which this condition is satisfied, and argue that it is associated to the $AdS_5$ non-Abelian T-dual solution presented in Section \ref{NATDKW}.


\subsection{IIA brane realization and M-theory uplift}

\begin{figure}[ht]
	\begin{center}
		\includegraphics[width=15cm]{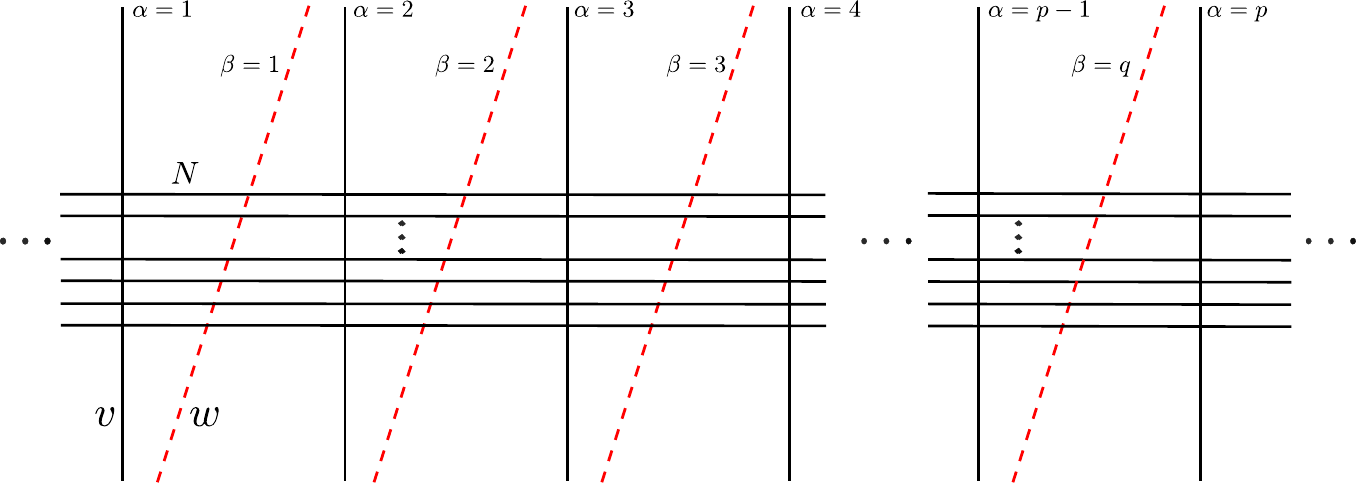}
	\end{center}
		\caption{The brane set-up associated to the Bah-Bobev $\mathcal{N}=1$ linear quivers. Vertical lines represent NS5-branes extended along $\{x_4,x_5\}$, denoted in \cite{Bah:2013aha} as $v$-branes, while diagonal lines represent the NS5'-branes extended along  $\{x_7,x_8\}$, denoted as $w$-branes. The same number of D4-branes extended along the $x_6$ direction stretch between adjacent 5-branes.}
		\label{BBbranesetup}
\end{figure}

Interestingly, it was shown in \cite{Bah:2013aha} that the linear quivers discussed above have a natural description in terms of D4, NS5, NS5' brane set-ups that generalize the $\mathcal{N}=2$ brane constructions in \cite{Witten:1997sc}, and allow for an M-theory interpretation. The two types of NS5-branes in this construction are taken to be orthogonal to each other, explicitly breaking $\mathcal{N}=2$ supersymmetry to $\mathcal{N}=1$. The specific locations of the branes involved are
\begin{itemize}
	\item $N$ coincident D4-branes extend along $\mathbb{R}^{1,3}$ and the $x_6$ direction. 
	\item $p$ non-coincident NS5-branes extend along $\mathbb{R}^{1,3}\times \{x_4,x_5\}$, and sit at $x_6=x_6^\alpha$ for $\alpha=1,\ldots,p$.  
	\item $q$ non-coincident NS5'-branes extend along $\mathbb{R}^{1,3}\times \{x_7,x_8\}$, and sit at $x_6=x_6^\beta$ for $\beta=1,\ldots,q$.  
\end{itemize}
The corresponding brane set-up is depicted in Figure  \ref{BBbranesetup}, see also \cite{Bah:2013aha}.

In this configuration, open strings connecting D4-branes stretched between two parallel NS5-branes are described at long distances and weak coupling by an $\mathcal{N}=2$ $SU(N)$ vector multiplet, while those connecting D4-branes stretched between perpendicular NS5 and NS5' branes are described by an $\mathcal{N}=1$ $SU(N)$ vector multiplet.  In turn, open strings connecting adjacent D4-branes separated by a NS5-brane (NS5'-brane) are described at low energies by bifundamental hypermultiplets with charge $\sigma_i=1$ ($\sigma_i=-1$). Finally, semi-infinite $N$ D4-branes (or D6 branes) at both ends of the configuration yield two sets of hypermultiplets in the fundamental representation of $SU(N)$. The resulting field theory is effectively four dimensional at low energies compared to the inverse size of the D4 along $x_6$. The effective gauge coupling behaves as $\frac{1}{g_4^2}\sim \frac{x_{6,n+1}-x_{6,n}}{g_s\sqrt{\alpha^\prime}}$. Given that the 5-branes can be freely moved along the $x_6$ direction, the gauge couplings are marginal parameters. 
Rotations in the $v=x_4+i x_5$ and $w=x_7+ i x_8$ planes of the NS5 and NS5' branes give a $U(1)_v$ and a $U(1)_w$ global symmetry, so that the IR fixed point R-symmetry and flavor $U(1)$ are realized geometrically as linear combinations of them:
\begin{equation}
R_0= U(1)_v+U(1)_w\, , \qquad {\cal F}=U(1)_v-U(1)_w\, .
\end{equation}


Relying on similar $\cN=2$ constructions in \cite{Witten:1997sc}, it is possible to describe the previous system of intersecting branes at strong coupling in M-theory. The $x_6$ direction is combined with the M-theory circle $x_{11}$ to form a complex coordinate $s=(x_6+ix_{11})/R_{11}$ describing a Riemann surface $\Sigma_2$, which is a punctured sphere or, equivalently, a punctured cylinder. The uplift of this system yields,

\begin{itemize}
	\item $N$ M5-branes wrapping the cylinder, from the $N$ D4-branes extended on $x_6$.
	\item $p$ simple punctures (in the language of \cite{Gaiotto:2009we}) on the cylinder, coming from the $p$ transversal M5-branes with flavor charge $\sigma_i=1$.
	\item $q$ simple punctures on the cylinder, coming from the $q$ transversal M5-branes with flavor charge $\sigma_i=-1$.
	\item Two maximal punctures, coming from the stacks of $N$ transversal M5-branes at both  ends of the cylinder. They are also assigned $\sigma_0,\sigma_\ell=\pm1$, from which the additional parameter $\k=(\sigma_0+\sigma_\ell)/2$ is defined, taking values $\k=-1,0,+1$.
\end{itemize}
The cylinder or sphere the M5-branes wrap can be viewed as a Riemann surface $\cC_{g,n}$ of genus $g=0$ and $n=p+q+2$ punctures, so that $\Sigma_2=\cC_{0,n}$. This Riemann surface can be deformed by bringing some of the punctures close to each other (which corresponds to certain weak and strong coupling limits of the dual 6d $\cN=(0,2)$ $A_{N-1}$ field theory living on the M5-branes) to a collection of higher-genus and less-punctured surfaces. The $\kappa$ parameter is associated to the type of punctures on the $\cC_{g,n}$ Riemann surface.

This closes our summary of the findings of the paper \cite{Bah:2013aha}, that we will use in the next section. Let us now propose a dual CFT to our background in eqs.(\ref{eq:metric_NATD})-(\ref{RRpotentials}).

\section{The non-Abelian T-dual of Klebanov-Witten as a $\mathcal{N}=1$ linear quiver} \label{quiverNATD}

As we showed in Section  \ref{quantized-charges}, the analysis of the quantized charges of the non-Abelian T-dual solution is consistent with a D4, NS5, NS5' brane set-up in which the number of D4-branes stretched between the NS5 and NS5' branes increases by $N_6$ units every time a NS5-brane is crossed. This configuration thus generalizes the brane set-ups discussed in the previous section and in \cite{Bah:2013aha}.  

In this section, inspired by the previous analysis, we will use the brane set-up depicted in Figure \ref{fig:NATDbranes} to propose a linear quiver dual  to the background in eqs.\! (\ref{eq:metric_NATD})-(\ref{RRpotentials}). As a consistency check we will compute its central charge using $a$-maximization and show that it is in perfect agreement with the holographic study in Section \ref{centralcharge} and the result of eq.(\ref{eq:cc_NATD_0n}), in particular. We will show that the central charge also satisfies the well-known 27/32 ratio  \cite{Tachikawa:2009tt} with the central charge associated to the non-Abelian T-dual of $AdS_5\times S^5/\mathbb{Z}_2$. This suggests defining our ${\cal N}=1$ conformal field theory as the result of deforming by mass terms  the $\mathcal{N}=2$ CFT associated to the non-Abelian T-dual of $AdS_5\times S^5/\mathbb{Z}_2$.

\subsection{Proposed $\mathcal{N}=1$ linear quiver}

The quantized charges associated to the non-Abelian T-dual solution are consistent with a brane set-up, depicted in Figure  \ref{fig:NATDbranes}, in which D4-branes extend on $\RR^{1,3}\times\{\rho\}$, NS5-branes on $\RR^{1,3}\times S^2(\theta_1,\phi_1)$ and NS5'-branes on $\RR^{1,3}\times {\tilde S}^2(\chi,\xi)$. This produces for $\rho\in [n\pi\alpha^\prime/L^2,(n+1)\pi\alpha^\prime/L^2]$, $n\rightarrow\infty$
and upon compactification, the  brane set-up, depicted in \autoref{fig:ATDbranes1} in Appendix \ref{appendixa}, associated to the Abelian limit of the solution. 

We conjecture that, in a similar fashion, the non-Abelian T-dual background in eqs. (\ref{eq:metric_NATD})-(\ref{RRpotentials}),   arises as the decoupling limit of a D4, NS5, NS5' brane intersection\footnote{Similar assumptions have been made in  \cite{Lozano:2016kum,Lozano:2016wrs}, with successful outcomes. 
}.  According to this proposal, we would have an {\it infinite-length } quiver  with (in the notation of Section \ref{subsec_linear_quivers}) $p=n$, $q=n$, $\ell=p+q=2n$ and  $z=(p-q)/\ell = 0$ with $n\rightarrow \infty$. The associated field theory would consist on $(2n-1)$ $\mathcal{N}=1$ vector multiplets and matter fields connecting them. However, this infinitely-long quiver {\it does not } describe a four dimensional field theory (its central charge is strictly infinite, among other problematic aspects). This is the same issue that we discussed when calculating the holographic central charge in Section \ref{centralcharge}. Some regularization is needed and, as we will see, the field theory precisely provides the way to do this.

\begin{figure}[ht]
	\begin{center}
			\includegraphics[width=15.5cm]{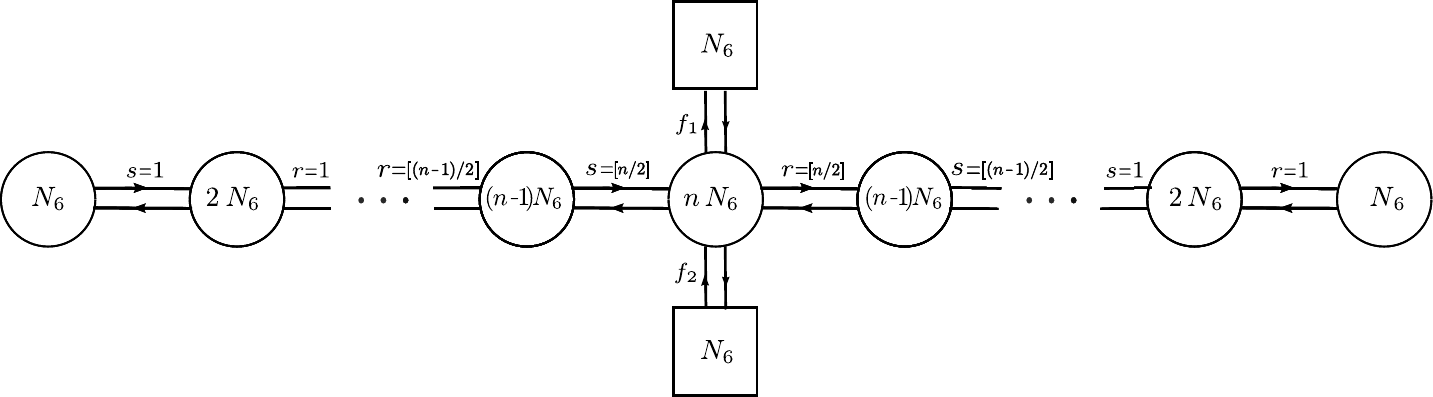}
	\end{center}
	\caption{Linear quiver proposed as dual to the non-Abelian T-dual solution. There are two matter fields $Q_j, \tilde Q_j$ in the bifundamental and anti-bifundamental of each pair of nodes, associated to a 5-brane connecting adjacent D4-stacks, with a total number of $j=1,\ldots,n-1$ hypermultiplets $H_j=(Q_j,\tilde Q_j)$ at each side of the quiver. We label $r=1,\ldots,\left[n/2\right]$ the $\sigma_j=+1$ hypermultiplets corresponding to NS5-branes and $s=1,2,\ldots,\left[n/2\right]$ the $\sigma_j=-1$ hypermultiplets from NS5'-branes, assuming an alternating distribution of both types of 5-branes.
This configuration comes from a re-ordering of the branes in Figure \ref{fig:NATDbranes} that is consistent with Seiberg self-duality and the vanishing of the beta functions and R-symmetry anomalies. 
The squares in the middle of the quiver denote flavor groups corresponding either to semi-infinite D4-branes ending on the NS5 and NS5' branes or to D6-branes transversal to the D4-branes. They {\it complete} the quiver at finite $n$. We choose $\sigma_{f_1}=-\sigma_{f_2}$ for the corresponding fundamental hypermultiplets.
	}
	\label{fig:NATDquiver}	
\end{figure}

Elaborating on the ideas in \cite{Lozano:2016kum}, we propose to study this quiver for finite $n$, {\it completing} it as shown in Figure \ref{fig:NATDquiver}.
The proposed field theory has the following characteristics:
\begin{itemize}
\item{It is strongly coupled. This is in correspondence with the fact  that it should be dual to an $AdS$ solution whose internal space is smooth in a large region and reduces to our non-Abelian T-dual background  in eqs. (\ref{eq:metric_NATD})-(\ref{RRpotentials}) in some limit.}
\item{The field theory is self-dual under Seiberg duality. This can be quickly seen, by observing that each node is at the self-dual point (with  $N_f=2N_c$).}
\item{The beta function and the R-symmetry anomalies vanish, in correspondence with the $SO(2,4)$ isometry of the background and the number of preserved SUSYs.}
\item{The central charge calculated by field theoretical means coincides (for long enough quivers) with the holographic result of eq.(\ref{eq:cc_NATD_0n}).}
\item{The quiver can be thought of as a mass deformation of the ${\cal N}=2$ quiver dual to the non-Abelian T-dual of   $AdS_5\times S^5/\mathbb{Z}_2$.} 
\end{itemize} 
 %
Below, we show that the field theory represented in Figure \ref{fig:NATDquiver} has all these characteristics. As it happens in the paper \cite{Lozano:2016kum}, the completion we propose with the flavor groups has the effect of ending the space at a given finite value in the $\rho$ direction. 

\subsection{$\beta$-functions and R-symmetry anomalies}

In this section we study  the $\beta$-functions and the anomalies associated to the linear quiver proposed in Figure \ref{fig:NATDquiver}. This analysis clarifies that the quantum field theory flows to a conformal fixed point in the infrared.

In a supersymmetric gauge theory, the $\beta$-function for a coupling constant $g$ is given by the 
well-known Novikov-Shifman-Vainshtein-Zakharov (NSVZ) formula \cite{Novikov:1983uc}, which can be written in terms of the number of colors, $N_c$, the number of flavors, $N_{f_q}$, and the anomalous dimensions for the matter fields, $\g_q$, as
\begin{equation}
\label{beta-function}
 \beta_g \sim 3 \, N_c - \sum\limits_{q} N_{f_q} \, (1 - \g_q) \ .
\end{equation}
Here, we considered the Wilsonian beta function. The denominator in the NSVZ formula is not relevant for us (see  \cite{ArkaniHamed:1997mj}
 for a nice explanation of this).
Another important quantity is the R-symmetry anomaly, given by the correlation function of three currents and represented by the Feynman diagram,
\begin{figure}[h]
	\centering
	\includegraphics[scale=0.4]{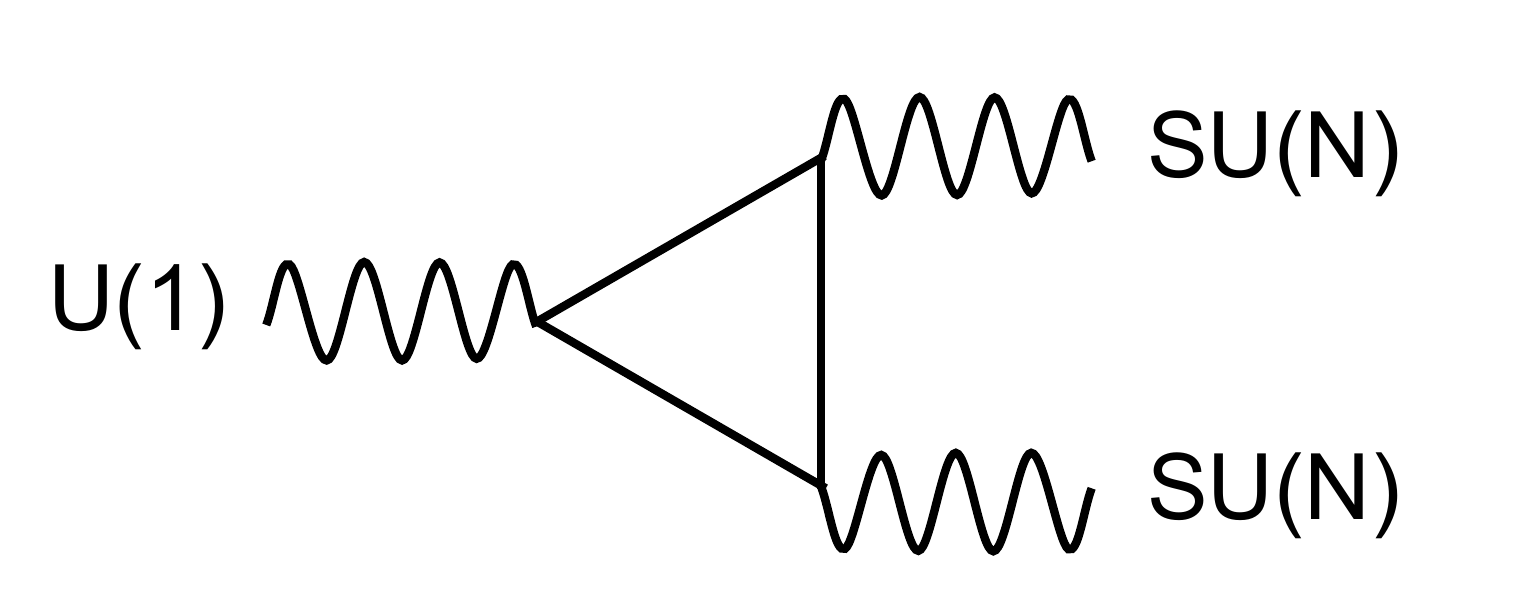}
	\caption{The R-symmetry anomaly.}
\label{Anomaly}
\end{figure}

\noindent The anomaly is given by the relation,
\begin{equation}
 \Braket{\partial^{\mu} J^5_{\mu}} \sim \D \Theta \, F \, \tilde{F} \ , \qquad \D \Theta = \sum R_f \, T(R_f) \ ,
\end{equation}
where $R_f$ is the R-charge of the fermions in the multiplet. In the case of an $SU(N)$ gauge group
\begin{equation}
 T(R_f) = \left\{ \begin{array}{ll}
 			 2N , & \textrm{for fermions in the adjoint representation}
			 \\[10pt]
			 1 ,   &  \textrm{for fermions in the fundamental representation}
 			\end{array} \right. 
\end{equation}
Moreover, at the conformal point, one should take into account the relation between the physical dimension of a gauge invariant operator  ${\cal O}$ (with engineering dimension $\D_{\cal O}$) and its R-charge $R_{\cal O}$,
\begin{equation}
  \textrm{dim}  ~{\cal O} = \D_{{\cal O}} + \frac{\g_{{\cal O}}}{2} = \frac{3}{2} \, R_{{\cal O}} \ .
\end{equation}
In the Appendix \ref{appendixb}, we present details of these calculations for the well-known example of the Klebanov-Witten CFT. Readers unfamiliar with that example can study the details in  Appendix \ref{appendixb} and then come back to the more demanding calculation presented  below.

Let us now analyze the quiver depicted in  Figure \ref{fig:NATDquiver}. We propose for the anomalous dimensions  and R-charges of the matter fields and gauginos the same values as in the  Klebanov-Witten CFT,
\begin{equation}
 \g_{Q} = \g_{\tilde{Q}} =  -\frac{1}{2} \ , R_Q=R_{\tilde{Q}}=\frac{1}{2},\;\;\; R(\lambda)=1.
\end{equation}
Notice that in our proposal only one bifundamental field runs in each arrow. We call them $Q$ or $\tilde{Q}$ depending on the direction of the arrow.
%
We find, substituting in eq. (\ref{beta-function}) for the nodes with rank $kN_6$,
\begin{equation}
\beta_k\sim 3 \, k \, N_6 -   \Big( (k+1) \, N_6 + (k-1) \,N_6)\Big)  \left(1  +\frac{1}{2}  \right)   = 0 \ , \qquad k = 1, \dots ,n \, .
\end{equation}
The first term reflects the contribution of the gauge multiplets and the second that of the matter fields. For the anomaly we find, 
\begin{equation}
\D \theta_k = 2 \, k \, N_6 + 2 \, \Big( (k+1) \, N_6 + (k-1) \, N_6 \Big)  \Big(\! - \frac{1}{2}  \Big) = 0 \ ,\qquad k = 1, \dots ,n \, .
\end{equation}
The first term indicates the contribution of the gauginos and the second one the contribution of the fermions in the $Q,\tilde{Q}$ multiplets.

These calculations indicate that both R-symmetry anomalies and beta functions are vanishing. Indeed, they belong to the same anomaly multiplet. Also, notice that the large anomalous dimensions indicate that the CFT is strongly coupled. With this numerology, we calculate that 
\begin{equation}
 \textrm{dim} ~Q= \textrm{dim} ~\tilde{Q}=\frac{3}{4}.
 \end{equation}
This allows for the presence of superpotential terms involving  four matter multiplets, like the ones proposed in \cite{Bah:2013aha}. Let us move now to the calculation of the central charge.

\subsection{Field-theoretical central charge}\label{sec:cc_compu}
In this section we compute the central charge of the quiver depicted in Figure \ref{fig:NATDquiver} at the fixed point, using the $a$-maximization procedure \cite{Intriligator:2003jj}. 

As recalled in Section \ref{subsec_linear_quivers}, the $a$ and $c$ central charges can be computed from
the $\cN=1$ R-symmetry t'Hooft anomalies of the fermionic degrees of freedom of the theory,
\begin{align}\label{eq:ac_charges}
 a(\epsilon)= \frac{3}{32} \, \big( 3 \, \Tr\,R_\epsilon^3 - \Tr\,R_\epsilon \big) \,,\;\;\;\; c(\epsilon)= \frac{1}{32} \, \big( 9 \, \Tr\,R_\epsilon^3 - 5 \, \Tr\,R_\epsilon \big).
\end{align}
The $R$-symmetry is given by $R_\epsilon=R_0+\frac{1}{2} \, \epsilon \, {\cal F}$.
Assigning charges $R_0(Q_j)=R_0(\tilde Q_j)=1/2$ to the chiral multiplet scalars, we have that
\begin{equation*}
 R_\epsilon(Q_j)= R_\epsilon(\tilde Q_j)= \frac{1}{2} \, \big( 1+ \epsilon\, \sigma_j \big) \,,
\end{equation*}
and, for the fermions
\begin{equation*}
R_\epsilon(\psi_j)=R_\epsilon(\tilde \psi_j)=  \frac{1}{2} \, \big( -1+ \epsilon \,\sigma_j \big) \,.
\end{equation*}
Now, we can compute the linear contribution to the anomaly coming from the hypermultiplet $H_j=(Q_j,\tilde Q_j)$, whose quiral fields transform in the fundamental of a gauge group with rank $N_a$ and in the anti-fundamental of another gauge group with rank $N_b$, and vice-versa:
\begin{equation}\label{eq:hypers}
 \Tr\,R_\epsilon(H_j) = N_a\,N_b \left(R_\epsilon(\psi_j)+R_\epsilon(\tilde\psi_j)\right)= N_a\,N_b(\epsilon\,\sigma_j-1)\, .
\end{equation}
The cubic contribution is 
\begin{equation}
\Tr\,R^3_\epsilon(H_j) = N_a\,N_b \left(R^3_\epsilon(\psi_j)+R^3_\epsilon(\tilde\psi_j)\right)=2 N_a\,N_b \left[\frac{1}{2}(\epsilon\,\sigma_j-1)\right]^3 .
\end{equation}

In turn, the linear and cubic anomaly contributions from an $\cN=1$ vector multiplet $V_t$ are given by,
\begin{equation}
\Tr\,R_\epsilon(V_t) = \Tr\,R^3_\epsilon(V_t) = N_a^2-1\,,
\end{equation}
where we have used  that  $R_\epsilon(\lambda)=R_0(\lambda)=1$ for the gaugino.

We now consider the completed quiver in Figure \ref{fig:NATDquiver}. Hypermultiplets with $\s_j=+1$ and $\s_j=-1$ (transforming in the bifundamental of gauge groups of ranks $N_j$, $N_{j+1}$) alternate along the quiver, and $\s_{f_1}=-\s_{f_2}$.  In this way all nodes are equipped with $\cN=1$ vector multiplets. Moreover, we have $z=0$ exactly, as well as $\kappa=0$. 
The total linear contribution of the hypermultiplers is then:
\begin{align}\label{eq:TrRH}
  \Tr\, R_\epsilon(H)
   &= \sum_{j=1,\text{left}}^{n-1} \Tr\, R_\epsilon(H_j) + \sum_{j=1,\text{right}}^{n-1} \Tr\, R_\epsilon(H_j) + \sum_{i=1}^2 \Tr\, R_\epsilon(H_{f_i}) \nonumber
   \\[5pt]
   &=N_6^2 \Bigg\{\sum_{j=1}^{n-1} j \, \big(j + 1\big) \big(\e\,\cancelto{0}{(\s_{j,\text{left}}+\s_{j,\text{right}})} -1\big) + n\,\sum_{i=1}^2 \big(\e\,\s_{f_i} - 1\big) \Bigg\} \nonumber
\end{align}
\begin{align}
   \phantom{\Tr\, R_\epsilon(H)}
   &=N_6^2 \Bigg\{ -2\sum_{j=1}^{n-1} j\,\big(j + 1\big) + n \big(\e\,\cancelto{0}{(\s_{f_1}+\s_{f_2})} - 2\big) \Bigg\} \nonumber
   \\[5pt]   
   & = N_6^2 \Bigg\{ -\frac{2}{3}\,n\big(n^2-1\big)  -2n \Bigg\} = N_6^2 \Bigg\{ -\frac{2}{3}n^3   -\frac{4}{3}n  \Bigg\} 
   \nonumber \\[5pt]
  &\approx -\frac{2}{3}n^3\,N_6^2 + \cO(n)\;. 
\end{align}
In the last line the approximation of a {\it long quiver} (large $n$) has been used. Similarly, the total cubic contribution of the hypermultiplets can be readily computed to be,
\begin{align}\label{eq:TrRH3}
  \Tr\, R^3_\epsilon(H)
   &= \sum_{j=1,\text{left}}^{n-1} \Tr\, R^3_\epsilon(H_j)  + \sum_{j=1,\text{right}}^{n-1} \Tr\, R^3_\epsilon(H_j) +  \sum_{i=1}^2 \Tr\, R^3_\epsilon(H_{f_i}) \nonumber
   \\[5pt]
   &=N_6^2 \Bigg[ \sum_{j=1}^{n-1} j \, \big(j + 1\big) \frac{1}{4}\Big( \big(\e\,\s_{j,\text{left}} -1\big)^3 + \big(\e\,\s_{j,\text{right}} -1\big)^3\Big)  + n\,N_6^2\sum_{i=1}^2 \frac14\big(\e\,\s_{f_i} - 1\big)^3 \Bigg] \nonumber
   \\[5pt]
   &=\frac{N_6^2}{4} \Bigg[ -2(1+3\e^2)\sum_{j=1}^{n-1} j\,\big(j + 1\big) + n\sum_{i=1}^2 \big(\e\,\s_{f_i} - 1\big)^3 \Bigg] \nonumber
 \\[5pt] 
   & = \frac{N_6^2}{12} \Bigg[  -2\big(1+3\e^2\big)n^3 -(12\e^2+4) n 
    \Bigg] \approx -\frac{1}{6} \, n^3 \, N_{6}^2 \, \big( 1 + 3 \, \epsilon^2 \big) + \cO(n),
\end{align}
where long quivers have been considered in the last expression. In turn, recalling that each node appears twice in the quiver depicted in Figure \ref{fig:NATDquiver}, with the exception of the central one, the trace anomaly coming from the $\cN=1$ vector multiplets becomes, 
\begin{align}
\Tr\, R_\epsilon(V) &= \Tr\, R^3_\epsilon(V) =2 \sum_{t=1}^{n-1} \Tr\, R_\epsilon(V_t) + \Tr\, R_\epsilon(V_n)= 2 \sum_{t=1}^{n-1} \big( t^2N_6^2-1 \big) + \big( n^2N_6^2-1 \big)
\nonumber\\
 &  =\frac{N_6^2}{3} \left(2n^3+n\right)-2(n -1) \approx \frac{2}{3} \, n^3\,N_{6}^2 + \cO(n)\;.
\end{align}
From this result we see that $\Tr\, R_\epsilon(V) \approx -\Tr\, R_\epsilon(H)$ in the large $n$ limit, so that the overall linear trace anomaly is of order $n\,N_{6}^2$ at most. Putting all these expressions together we find, for the exact charges in eq.~\eqref{eq:ac_charges},
\begin{align}
 a(\e) 
 &= \frac{3N_6^2}{64}\Bigg\{  3(1-\e^2)n^3 +  2(1 - 3\e^2 )n - \frac{4}{N_6^2}(2n-1)  \Bigg\} , \nonumber 
 \\[5pt]
 c(\e) 
 &= \frac{N_6^2}{64}\Bigg\{  9(1-\e^2)n^3 + 2(5-9\e^2)n - \frac{8}{N_6^2}(2n-1) 
 \Bigg\} .  
\end{align}
From these expressions we see that $a(\e)$ is clearly maximized for $\e=0$, as expected for the $\cN=1$ fixed point \cite{Bah:2013aha}. The superconformal central charges are thus found to be
\vspace{-2mm}
\begin{align}\label{eq:exact_N1_ac}
a_{\cN=1} \equiv a(\e=0) &= \frac{3}{64}\left\{  (3n^3 + 2n)N_6^2 -4(2n-1) \right\} , \nonumber \\[7pt]
c_{\cN=1} \equiv c(\e=0) &=\frac{1}{64}\left\{  (9n^3 +  10n)N_6^2 -8(2n-1) \right\}.
\end{align} 
They give, in the large $n$ limit,
\begin{equation}
\label{centralchargeFT}
  c_{\cN=1} \approx a_{\cN=1} \approx \frac{9}{64} \, n^3 \, N_{6}^2   + \cO(n) \,.
\end{equation} 
This final result matches  the holographic calculation given by eq. \eqref{eq:cc_NATD_0n}. This provides  a non-trivial check of the validity of the linear quiver in Figure \ref{fig:NATDquiver} as dual to the background in eqs. (\ref{eq:metric_NATD})-(\ref{RRpotentials}). It is noteworthy that the agreement with the holographic result occurs in the large number of nodes limit, $n\rightarrow\infty$. 

\begin{figure}[h]
	\centering
	\includegraphics[scale=1.5]{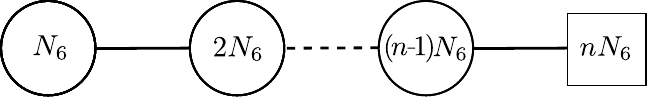}
	\caption{Completed quiver associated to the non-Abelian T-dual of $AdS_5\times S^5$. Each line represents a hypermultiplet of ${\cal N}=2$ SUSY.}
\label{Sfetsos-Thompson-quiver}
\end{figure}

A further non-trivial check of the validity of our proposed quiver is that the central charge given by (\ref{centralchargeFT}) and that associated with the non-Abelian T-dual of $AdS_5\times S^5/\mathbb{Z}_2$ satisfy the same {\it 27/32 relation} \cite{Tachikawa:2009tt}, that is, 
\begin{equation}
\label{27/32}
c_{\cN=1}= \frac{27}{32}\,c_{\cN=2}, 
\end{equation}
as the  central charges of the corresponding theories prior to dualization. Indeed, the quiver associated to the non-Abelian T-dual of $AdS_5\times S^5/\mathbb{Z}_2$ can be obtained by modding out by $\mathbb{Z}_2$ the quiver describing the non-Abelian T-dual of 
$AdS_5\times S^5$, constructed in \cite{Lozano:2016kum} and depicted in Figure \ref{Sfetsos-Thompson-quiver}. This quiver was completed at finite $n$ by a flavor group with gauge group $SU(nN_6)$. It thus satisfies the condition to be conformal (preserving ${\cal N}=2$ SUSY), i.e. that the number of flavors is twice the number of colors at each node. 
Modding out by $\mathbb{Z}_2$ results in the same quiver in Figure  \ref{fig:NATDquiver}, but built out of $2n$ $\mathcal{N}=2$ vector and matter multiplets. 
Taking the central charge, computed in \cite{Lozano:2016kum}, for the non-Abelian T-dual of $AdS_5\times S^5$ and doubling it, 
we obtain the central charge of the non-Abelian T-dual of $AdS_5\times S^5/\mathbb{Z}_2$
\begin{equation}
c_{NATD\, AdS_5\times S^5/\mathbb{Z}_2} \approx 2\times \frac{1}{12} \, n^3 N_{6}^2+\cO \big(n \big),\label{papa}
\end{equation}
and we find that eq. (\ref{27/32}) indeed holds with $c_{\mathcal{N}=1}$ as in eq. (\ref{centralchargeFT}) and $c_{\mathcal{N}=2}$ as in eq. (\ref{papa}).
We have checked in Appendix \ref{centralchargeN=2} that the same result (\ref{papa}) is reproduced 
using $a$-maximization.
 The $a$-charge is maximized for $\epsilon=\frac{1}{3}$, as previously encountered in  \cite{Bah:2013aha}.

Further, one can check that also at finite $n$, $a_{\mathcal{N}=1}$ and $c_{\mathcal{N}=1}$ satisfy the relation \cite{Tachikawa:2009tt} \footnote{We would like to thank Nikolay Bobev for suggesting this to us.},
\begin{equation}\label{eq:TW}
  a_{\cN=1} = \frac{9}{32}\big(4\,a_{\cN=2}- c_{\cN=2}\big) \,, \qquad   c_{\cN=1} = \frac{1}{32}\big(-12\,a_{\cN=2} + 39\,c_{\cN=2}\big)
\end{equation}
with the $a_{\mathcal{N}=2}$, $c_{\mathcal{N}=2}$ exact central charges of the $\mathcal{N}=2$  quiver. The explicit expressions for $a_{\mathcal{N}=2}$ and $c_{\mathcal{N}=2}$ are given in eq. (\ref{eq:exact_N2_ac}) in Appendix  \ref{centralchargeN=2}. This precisely defines our dual CFT as the result of deforming by mass terms the CFT dual to the Sfetsos-Thompson solution modded by $\mathbb{Z}_2$.

 The material presented in this section makes very precise
 the somewhat  loose ideas proposed in the works \cite{Itsios:2013wd}. In particular, we have identified the concrete relation via a RG-flow between the non-Abelian T-dual of $AdS_5\times S^2/\mathbb{Z}_2$ and the non-Abelian T-dual of the Klebanov-Witten solution. Notice that here, we are providing  precisions about the CFT dual to the non-Abelian T-dual backgrounds. This more precise information is matched by the regularized form of the non-Abelian T-dual solution.
 
The  diagram in the Introduction section  summarizes the connections between the UV and IR field theories discussed in this section. We repeat it here for the perusal of the reader.

\begin{equation*}
 \xymatrixrowsep{4pc}
 \xymatrix{ 
         AdS_5 \times S^5/\mathbb{Z}_2  \ar[d]^{\textrm{mass}}  \ar[rr] && \textrm{ NATD of } AdS_5 \times S^5/\mathbb{Z}_2\ar[d]^{\textrm{mass}} \\
       AdS_5 \times T^{1,1} \ar[rr] &&  \textrm{NATD of } AdS_5 \times T^{1,1}
           }
\end{equation*}
As a closing remark, an explicit flow (triggered by a VEV) between the ${\cal N}=1$ and the ${\cal N}=2$ non-Abelian T-dual backgrounds was constructed in
\cite{Macpherson:2015tka}. It should be interesting to use the detailed field theoretical picture developed above and 
in \cite{Lozano:2016kum}, to be more precise about various aspects of this RG-flow.

\section{Solving the INST-BBBW puzzle}
\label{section5}

The non-Abelian T-dual of the Klebanov-Witten background was first written in \cite{Itsios:2013wd} (INST).  Further, in that paper an attempt was made to match 
the non-Abelian T-dual background with a Bah, Beem, Bobev and Wecht (BBBW) solution \cite{Bah:2012dg}.
This matching was feasible assuming a particular split of the metric into a seven-dimensional and a four-dimensional internal space (see below). The formula in \cite{Bah:2012dg} for the central charge of BBBW solutions led however to  $a\sim c\sim 0+ O(N)$ for the non-Abelian T-dual solution, in blatant disagreement with the holographic result.
This was the puzzle that the authors of \cite{Itsios:2013wd} pointed out. 
In this section we present its resolution. We start by summarizing the most relevant aspects of the work \cite{Bah:2012dg}.

In the work of Bah, Beem, Bobev and Wecht new $\mathcal{N}=1$ $AdS_5$ solutions in M-theory were constructed, describing the fixed points of new  $\mathcal{N}=1$ field theories associated to M5-branes wrapped on complex curves. The central charges of these SCFTs were computed using the six dimensional anomaly polynomial and $a$-maximization, and were shown to match, in the large number of M5-branes limit, the holographic results.

The solutions constructed in \cite{Bah:2012dg} were obtained by considering
M-theory compactified  on
a deformed four-sphere. In principle, this compactification leads to an $SO(5)$-gauged supergravity in seven dimensions. Following the ideas in \cite{Maldacena:2000mw}, BBBW  searched for their solutions in the seven dimensional gravity theory (a $U(1)^2$ truncation of the full $SO(5)$ theory) discussed in \cite{Cvetic:1999xp}. They proposed a background consisting of a metric, two gauge fields $A_\mu^{(i)}$ and two scalars $\lambda^{(i)}$, of the form
\begin{eqnarray}
\label{7d}
& & ds_7^2 = 
e^{2 f(r)}[dx_{1,3}^2 + dr^2] +e^{2g(r)} d\Sigma_k(x_1,x_2),\nonumber\\
& & F^{(1)}= \frac{p}{8g-8} vol \Sigma_k,\;\;\; F^{(2)}=\frac{q}{8g-8}  vol\Sigma_k,\;\;\; \lambda^{(i)}(r).
\end{eqnarray}
They then searched for `fixed point' solutions, namely, those 
where $\frac{d}{dr}\lambda^{(i)}=\frac{d}{dr}g=0$ and 
$f\sim -\log r$, leading to backgrounds of the form 
$AdS_5\times \Sigma_k$.
They found general solutions depending on four parameters $(N,\kappa,z,g)$. For excitations with wavelength longer than the size of $\Sigma_k$, 
these are dual to four dimensional CFTs. In the dual CFT $N$ is the number of M5-branes, $\kappa=\pm 1,0$ is the curvature of the 2d Riemann surface that they are wrapping, $g$ is its genus and $z$ is the so-called `twisting parameter', defined as $z=\frac{(p-q)}{2(g-1)}$ from the integer numbers $p,q$ that indicate the twisting applied to the M5-branes. The holographic central charge computed in 
\cite{Bah:2012dg} depends on these parameters, and reads
\begin{equation}
c=a=N^3(1-g)\Big[\frac{1-9z^2 +\kappa (1+3z^2)^{3/2}}{48 z^2}\Big].
\label{cc7d}
\end{equation}
BBBW completed their analysis deriving various of their formulas, in particular the holographic central charge,
using purely 4d CFT arguments. Their CFTs are combinations 
of Gaiotto's $T_N$-theories, conveniently gauged and connected with other $T_N$ factors, with either ${\cal N}=1$
or ${\cal N}=2$ vector multiplets (shaded and unshaded $T_N$'s in the same line as what we explained in Section \ref{subsec_linear_quivers}). 

The key-point to be kept in mind after this discussion is that these results were 
obtained in the context of a {\it compactification} of eleven-dimensional 
supergravity to seven dimensions. 

Let us now come back to the paper \cite{Itsios:2013wd}. The matching of the non-Abelian T-dual solution with a BBBW geometry assumed that the seven dimensional part of the metric in (\ref{7d})
was  $AdS_5\times S^2 (\theta_1,\phi_1)$ and 
 that the internal space contained the coordinates 
$[\rho,\chi,\xi,x_{11}]$. 
Also, the authors of \cite{Itsios:2013wd} chose the parameters $\kappa=z=1$
for such matching. Using the formula (\ref{cc7d}) in BBBW for the central charge they then found that at leading order the central charge vanished. 

What was not-correct in the analysis of \cite{Itsios:2013wd} was the assumption that the non-Abelian T-dual solution could be obtained from a {\it compactification} of M-theory on a deformed four-sphere
(and hence be in the BBBW class of solutions). In fact, inspecting the BPS equations of BBBW -- eq.\! (3.10) 
of \cite{Bah:2012dg} -- one finds that a fixed point solution 
does not exist for the set of values $\kappa=|z|=1$. 
Even more, the generic solution that BBBW wrote in their eq.\! (3.8) is troublesome for those same values. 

A parallel argument can be made by comparing the BBBW and 
non-Abelian T-dual solutions in the language of the paper 
\cite{Apruzzi:2015zna}.
Indeed, the comparison in the Appendix C of 
\cite{Apruzzi:2015zna},
 shows that 
these solutions fit in their formalism in Section 4.2 
 for values of
parameters that are incompatible. Either BBBW is 
fit {\it or} the non-Abelian T-dual solution is, 
for a chosen set of parameters.

The resolution to this problem is that
 the non-Abelian T-dual 
background should instead be thought of as providing a 
{\it non-compactification} of eleven dimensional supergravity. 
Strictly speaking, our coordinate $\rho$ 
runs in $[0,\infty]$, the four manifold is non-compact. 
In our calculation of the central charge, 
we {\it assumed} that the $\rho$-coordinate 
was bounded in $[0,n\pi\frac{\alpha'}{L^2}]$, 
but this hard cut-off, as we emphasized,  is not a geometrically satisfactory 
way of bounding a  coordinate. There should 
be another, more general solution, that contains our non-Abelian T-dual metric in a small patch of the space (for small values of $\rho$), and closes the $\rho$-coordinate at some large value $\rho_n=n\pi \frac{\alpha'}{L^2}$. But this putative new metric, especially its behaviour near $\rho_n$, will differ considerably from the one obtained via non-Abelian T-duality. {Below, we will comment more about this putative solution. }

Let us close with some field theoretical remarks. The class of CFTs studied by BBBW \cite{Bah:2012dg} are quite different from those studied by Bah and Bobev in \cite{Bah:2013aha}.
Their central charges are different, and the first involve Gaiotto's $T_N$ theories while the second do not. In the same line,  our CFT discussed in Section \ref{quiverNATD} is a generalization, but strictly different, of the theories in
\cite{Bah:2013aha}, and is certainly different from those in \cite{Bah:2012dg}. 

{The quiver we presented in Section \ref{quiverNATD} encodes the dynamics of a solution in Type IIA/M-theory where the $\rho$-coordinate is bounded in a geometrically sounding fashion. The addition of the flavor groups in our quiver encode the way in which the $\rho$-coordinate should be ended. Indeed, in analogy with what was observed in \cite{Lozano:2016kum,Lozano:2016wrs}, we expect the metric behaving like that of D6 branes close to the end of the space. In M-theory language, we expect to find a puncture on the Riemann surface, representing the presence of flavor groups in the dual CFT. We will be slightly more precise about this in the Conclusions section.}


\section{Conclusions and future directions}\label{conclusions}
Let us briefly summarize the main  achievements of this paper.

After discussing details of the Type IIA solution obtained by non-Abelian T-duality applied on the Klebanov-Witten background, we carefully studied its quantized charges
and holographic central charge (Section \ref{NATDKW}). We lifted the solution to M-theory and showed by explicit calculation of the relevant differential forms that the background has $SU(2)$-structure and fits the classification of \cite{Gauntlett:2004zh}.

Based on the quantized charges, we proposed a brane set-up (Section \ref{quiverNATD}) and a precise quiver gauge theory, generalizing the class of theories discussed by Bah and Bobev in \cite{Bah:2013aha} (and summarized in our Section \ref{subsec_linear_quivers}). This quiver was used to calculate the central charge,  one of the important observables 
of a conformal field theory at strong coupling. Indeed, in Section \ref{quiverNATD}, we showed the precise agreement of this observable, computed by field theoretical means, with the holographic central charge. We also showed that the quiver has a strongly coupled IR-fixed point. Finally, Section \ref{section5}, solves a puzzle raised in previous bibliography. Various appendices discuss technical points in detail. In particular, relations of the non-Abelian T-dual of the Klebanov-Witten background and the more conventional T-dual, details about the dual field theory, etc, are carefully explained there. 

To close this paper let us state  the most obvious and natural continuation of our work.
As we discussed, the holographic central charge calculation in Section \ref{NATDKW} was done for a {\it regulated} version of the Type IIA background. Indeed, the integral over the internal space was taken to range in a finite interval for the $\rho$-coordinate. We introduced a hard-cutoff, but emphasized that this form  of regularization is not rigorous from a geometric viewpoint. Fortunately, the dual CFT provides a rationale to regulate the space. The flavor groups $SU(N_6)$ that end our quiver field theory (see Figure
\ref{fig:NATDquiver}), will be reflected in the 
Type IIA background by the presence of flavor branes that will backreact and end the geometry, solving the Einstein's equations. In eleven dimensions, the same effect will be captured by punctures on the $S^2$ that the M5 branes are wrapping. A phenomenon like this was at work in the papers \cite{Lozano:2016kum,Lozano:2016wrs}.

The formalism to backreact these flavor D6 branes is far-less straightforward in the present case, as the number of isometries and SUSY is less than in the cases of \cite{Lozano:2016kum,Lozano:2016wrs}. Qualitatively one may think of defining the completed solution by deforming with mass terms the superposition of $\mathcal{N}=2$ Maldacena-Nunez solutions \cite{Maldacena:2000mw} used in  \cite{Lozano:2016kum} to complete the Sfetsos-Thompson background. This would give rise to a superposition of $\mathcal{N}=1$ MN solutions defining the completed non-Abelian T-dual solution. It is unclear however in which precise way this superposition would solve the (very non-linear) PDEs associated to $\mathcal{N}=1$ solutions \cite{Gauntlett:2004zh,Bah:2015fwa}. 
We see two possible paths to follow:
\begin{itemize}
\item{In the paper \cite{Bah:2015fwa}, Bah rewrote the general M-theory background of \cite{Gauntlett:2004zh} in terms of a new set of coordinates that are more useful to discuss the addition of  punctures on the Riemann surface. In the type IIA language the new solutions found using Bah's non-linear and coupled PDEs should represent the addition of the flavor  D6 branes argued above. The equations need to be solved close to the singularity (the puncture or the flavor D6 brane) and then numerically matched with the rest of the non-Abelian T-dual background.}
\item{In  \cite{Apruzzi:2015zna} generic backgrounds in {\it massive} Type IIA were found with an $AdS_5$ factor in the metric and preserving eight SUSYs. For the particular case in which the internal space contains a Riemann surface of constant curvature, the involved set of non-linear and coupled PDEs simplifies considerably. One of the solutions, for the case in which the massive parameter vanishes, is the one studied in this paper---named INST in \cite{Apruzzi:2015zna}. Since the paper \cite{Apruzzi:2015zna} and some follow-up works have discussed ways of ending these spaces by the addition of D6 and D8 branes, we could consider these technical developments together with the ideas discussed above.  }
\end{itemize}
Finding a {\it completed} or {\it regularized} solution would provide the first example 
for a background dual to a CFT like that discussed in Section \ref{quiverNATD}.  The natural following steps would be to extend the formalism to discuss the situations for a cascading QFT. In fact, the precise knowledge of the CFT we have achieved in this paper can be used to improve the understanding and cure the singularity structure of the backgrounds written in the first paper in \cite{Itsios:2013wd}, in \cite{bbranch}, etc. We reserve  these problems to be discussed in forthcoming publications.

\subsection*{Acknowledgements}
We would like to thank Fabio Apruzzi, Antoine Bourget, Noppadol Mekareeya, Achilleas Passias, Alessandro Tomasiello, Daniel Thompson, Salom\'on Zacar\'{\i}as and especially Nikolay Bobev and Niall Macpherson for very useful discussions. 
G.I. is supported by the FAPESP grants 2016/08972-0 and 2014/18634-9.
Y.L. and J.M. are partially supported by the Spanish and Regional Government Research Grants FPA2015-63667-P and
FC-15-GRUPIN-14-108. J.M. is supported by the FPI grant BES-2013-064815 of the Spanish MINECO, and the travel grant EEBB-I-17-12390 of the same institution. Y.L. would like to thank the Physics Department of Swansea U. and the Mainz Institute for Theoretical Physics (MITP) for the warm hospitality and support. J.M. is grateful to the Physics Department of Milano-Bicocca U. for the warm hospitality and exceptional working atmosphere.
 C.N. is Wolfson Fellow of the Royal Society.


\appendix

\section{Connection with the GMSW classification}
\label{appendixgmsw}

In this appendix we prove that the uplift of the non-Abelian T-dual of the Klebanov-Witten solution fits in the classification of $\mathcal{N}=1$ $AdS_5$ backgrounds in M-theory of GMSW \cite{Gauntlett:2004zh}. 

\subsection{Uplift of the non-Abelian T-dual solution }
\label{UpliftNATDKW}

The eleven dimensional uplift of the non-Abelian T-dual solution consists of metric and 4-form flux. The metric is given by
\begin{equation}
\label{MetricKW11}
 ds^2_{11} = e^{- \frac{2 \, \Phi}{3}} \, ds^2_{IIA} + e^{\frac{4 \, \Phi}{3}} \, \big(   dx_{11} + C_1  \big)^2 \ ,
\end{equation}
where $x_{11}$ stands for the $11^{th}$ coordinate, $ds^2_{IIA}$ is the ten dimensional metric, given by 
eq. (\ref{eq:metric_NATD}), 
$\Phi$ is the dilaton, given by eq. (\ref{eq:NATD_dilaton}), and $C_1$ is the RR potential given in eq. \eqref{RRpotentials}. The eleven dimensional four-form field, $F^M_4$, is derived from $F^M_4 = dC^M_3$, where
\begin{equation}
 C^M_3 = C_3 + B_2 \wedge dx_{11} \ ,
\end{equation}
and $C_3$ and $B_2$ are given by  \eqref{RRpotentials} and (\ref{eq:B2_NATD}), respectively.
The final expression for $F^M_4$ is given by
\begin{equation}
\label{F4KW11}
 \begin{aligned}
  F^M_4 & = - \frac{\a' \, \l^2 \, \big(   L^4 \, \l^4_1 + \a'^2 \, \r^2  \big)}{Q} \, \cos\chi \, d\Omega_2(\th_1, \phi_1) \wedge d\r \wedge dx_{11}
  \\[10pt]
  & + \frac{\a' \, L^4 \, \l^4_1 \, \l^2}{Q} \, \r \, \sin\chi \, d\Omega_2(\th_1, \phi_1)  \wedge d\chi \wedge dx_{11}
  \\[10pt]
  & + \frac{4 \, \a'^{3/2} \, L^4 \, \l^4_1 \, \l \, \big(   \l^2 - \l^2_1  \big)}{g_s \, Q} \, \r^2 \, \cos\chi \, \sin^2\chi \, d\Omega_2(\th_1, \phi_1) \wedge d\r \wedge d\xi  
\\[10pt]
  & + \frac{4 \, \a'^{3/2} \, L^4 \, \l^4_1 \, \l \, P}{g_s \, Q} \, \r^3 \, d\Omega_2(\th_1, \phi_1)  \wedge d\Omega_2(\chi, \xi) 
  \\[10pt]
  & + \frac{4 \, \a'^{3/2} \, L^4 \, \l^4_1 \, \l^3 \, \mathcal{S}}{Q^2} \, \r^2 \, \cos\th_1 \, d\Omega_2(\chi, \xi)  \wedge d\r \wedge d\phi_1
\\[10pt]
  & + \frac{\a'^3 \, \l^2 \, \mathcal{S}}{Q^2} \, \r^2 \, \sin\chi \, \big( d\xi + \cos\th_1  \, d\phi_1 \big) \wedge d\r \wedge d\chi \wedge dx_{11} \ ,
 \end{aligned}
\end{equation}
where for the sake of clarity we have defined,
\begin{equation}
 \mathcal{S} \equiv  Q -  2 \, L^4 \, \l^4_1 \, \big(   \l^2 + \l^2_1  \big) - 2 \, \a'^2 \, \r^2 \, \l^2_1 \ .
\end{equation}
In the large $n$ limit this expression takes the simpler form ($\l^2=1/9$, $\l_1^2=1/6$, $\a'=g_s=1$),
\beq
F_4^M \approx \frac{L^4}{27} (\rho-n\pi)\,  d\Omega_2(\th_1, \phi_1)  \wedge d\Omega_2(\chi, \xi) \ ,
\eeq
which tells us that the M5-branes sourcing this flux are transversal to both squashed two-spheres $S^2(\th_1, \phi_1)$ and $\tilde{S}^2(\chi, \xi)$. These are associated to the global isometries $U(1)_w$ and $U(1)_v$, whose product lies in the Cartan of both the local R-symmetry and the non-anomalous flavor symmetry.





\subsection{Review of GMSW}

Before matching the previous solution within the classification in \cite{Gauntlett:2004zh}, let us briefly review the most general $\cN = 1$ eleven dimensional solutions with an $AdS_5$ factor found in that paper. These solutions are described by a metric of the form,
\begin{equation}
\label{MetricGMSW}
 ds^2_{11} =e^{2 \, \L} \, \Big[  ds^2_{AdS_5} + ds^2_{M_4} + \frac{e^{- 6 \, \L}}{\cos^2 \zeta} \, dy^2 + \frac{\cos^2 \zeta}{9 \, m^2} \, \big(   d\tilde{\psi} + \tilde{\r}  \big)^2   \Big] \ .
\end{equation}
Here $\tilde{\psi}$ is the R-symmetry direction, $\tilde{\r}$ is a one-form defined on $M_4$, whose components depend on both the $M_4$ coordinates and $y$, and  $\L$ and $\zeta$ are functions also depending on the $M_4$ coordinates and $y$. The coordinate $y$ is related to the warping factor $\L$ and the function $\zeta$ through,
\begin{equation}
 2 \, m \, y = e^{3 \, \L} \, \sin \zeta \ ,
\end{equation}
with $m$ being the inverse radius of $AdS_5$.

The four-dimensional manifold $M_4$ admits an $SU(2)$ structure which is characterized by a $(1,1)$-form $J$ and a complex $(2,0)$-form $\Om$. The $SU(2)$ structure forms, together with the frame components $K^1$ and $K^2$, defined as,
\begin{equation}
 K^1 \equiv \frac{e^{- 3 \, \L}}{\cos \zeta} \, dy \, ,\qquad K^2 \equiv \frac{\cos \zeta}{ 3 \, m} \, \big(   d\tilde{\psi} + \tilde{\r}  \big) \, ,
\end{equation}
must satisfy the following set of differential conditions dictated by supersymmetry,
\begin{eqnarray}
 e^{-3 \, \L} \, d \big(  e^{3 \, \L} \, \sin \zeta  \big) & = & 2 \, m \, \cos \zeta \, K^1 \ , \label{DC1}
 \\[5pt]
 e^{-6 \, \L} \, d \big(  e^{6 \, \L} \, \cos \zeta \, \Om  \big) & = & 3 \, m \, \Om \wedge \big( - \sin \zeta \, K^1 + i \, K^2   \big) \ , \label{DC2}
\\[5pt]  
 e^{-6 \, \L} \,d  \big(  e^{6 \, \L} \, \cos \zeta \, K^2  \big) & = & e^{- 3 \, \L} \star G + 4 \, m \, \big(  J - \sin \zeta \, K^1 \wedge K^2  \big) \ , \label{DC3}
\\[5pt]
 e^{-6 \, \L} \, d \big(  e^{6 \, \L} \, \cos \zeta \, J \wedge K^2  \big) & = & e^{- 3 \, \L} \, \sin\zeta \, G + m \, \big(  J \wedge J - 2 \, \sin \zeta \, J \wedge K^1 \wedge K^2  \big) \ . \label{DC4}
\end{eqnarray}
In the above formulas, $\star$ stands for Hodge duality in the six-dimensional space spanned by $M_4$ and the one-forms $K^1$ and $K^2$. $G$ is an eleven-dimensional four-form whose components lie along the six-dimensional space that is transverse to $AdS_5$
\footnote{
There is a sign difference between the first term in the second line of \eqref{4formG} and the corresponding term in eq. (2.50) of \cite{Gauntlett:2004zh},  that is due to our different conventions for Hodge duality.
}
\begin{equation}
\label{4formG}
 \begin{aligned}
  G = & - \partial_y e^{- 6 \, \L} \, \widehat{\textrm{vol}}_4 - \frac{e^{-9 \, \L}}{\cos \zeta} \, \big(   \hat{\star}_4 d_4 e^{6 \, \L}  \big) \wedge K^1 - \frac{\cos^3 \zeta}{3 \, m} \, \big(   \hat{\star}_4 \partial_y \tilde{\r}  \big) \wedge K^2
  \\[5pt]
     & - \Big[  \frac{e^{3 \, \L}}{3 \, m} \, \cos^2\zeta \, \hat{\star}_4 d_4 \tilde{\r} + 4 \, m \, e^{-3 \, \L} \, \hat{J}   \Big] \wedge K^1 \wedge K^2 \ .
 \end{aligned}
\end{equation}
In this expression the hatted quantities are referred to the four-dimensional metric $\hat{g}^{(4)}_{\mu \nu} = e^{6 \, \L} \, g^{(4)}_{\mu \nu}$. Finally, $d_4$ is the exterior derivative on the four-dimensional space that is transverse to $AdS_5$ and $K^1 , \, K^2$.

\subsection{Recovering the non-Abelian T-dual from GMSW}

Let us now find the explicit map between the GMSW geometry and the lifted non-Abelian T-dual geometry. In order to do this we first identify the functions $\L$ and $\zeta$ according to,
\begin{equation}
 e^{6 \, \L} = 4 \, y^2 + \frac{q}{9} \ , \qquad \cos \zeta = \sqrt{\frac{q}{36 \, y^2 + q}} \ ,
\end{equation}
where $q$ is a function of the coordinates on $M_4$ and $y$, determined below. 
We also take
\begin{equation}
 \tilde{\r} = - \frac{1}{6 \, q} \, d w -\frac{1 - 12 \, q}{12 \, q} \, \cos \th_1 \, d \phi_1 \ .
\end{equation}
Then the  one-forms $K^1$ and $K^2$ read,
\begin{equation}
 K^1 = \frac{3}{\sqrt{q}} \, dy \ , \qquad K^2 = \frac{1}{3 \, m} \, \sqrt{\frac{q}{36 \, y^2 + q}} \, \Big[   d \tilde{\psi} - \frac{1}{6 \, q} \, d w -\frac{1 - 12 \, q}{12 \, q} \, \cos \th_1 \, d \phi_1  \Big] \ .
\end{equation}
Moreover, we define an orthogonal frame for the four-dimensional space $M_4$,
\begin{equation}
 \begin{aligned}
  e^1 & = \frac{1}{\sqrt{6}} \, \sin \th_1 \, d\phi_1 \ , \qquad e^2 = \frac{1}{\sqrt{6}} \, d\th_1
  \\[10pt]
  e^3 & = \frac{1}{18} \, \sqrt{\frac{36 \, q - 1}{36 \, y^2 \, q + q^2}} \, dz \ , \qquad e^4 = \frac{1}{18} \, \sqrt{\frac{36 \, q - 1}{36 \, y^2 \, q + q^2}} \, \Big(   dw + \frac{1}{2} \, \cos \th_1 \, d\phi_1  \Big) \ .
 \end{aligned}
\end{equation}
In the above expressions, $q$ can be thought of as a function of $z$ and $y$ through the relation,
\begin{equation}
 z -162 \, y^2 - \frac{36 \, q - 1}{12} - \frac{1}{12} \, \ln \big(   36 \, q - 1 \big) = 0 \ .
\end{equation}
Solving this equation for $q$ one finds\footnote{
With $\textrm{ProductLog}({\cal Z})$ we mean the solution of the equation ${\cal Z} = {\cal W} \, e^{\cal W}$ in terms of ${\cal W}$.
},
\begin{equation}
 q = \frac{1}{36} \, \Big[  1 + \textrm{ProductLog} \Big(   e^{12 \, (z - 162 \, y^2)}  \Big)   \Big] \ .
\end{equation}
From the above frame one can construct the forms $J$ and $\Om$ of the $SU(2)$ structure on $M_4$ as,
\begin{equation}
 J = e^1 \wedge e^2 + e^3 \wedge e^4 \ , \qquad \Om = e^{i \, \tilde{\psi}} \, \big(    e^1 + i \, e^2  \big) \wedge \big(   e^3 + i \, e^4 \big) \ .
\end{equation}
Both the metric and the 4-form flux associated to our solution are then obtained
after identifying\footnote{
We take $L = m = \a' = g_s = 1$ for convenience. There is a minus overall sign between $G$, from \eqref{4formG}, and $F_4$, from \eqref{F4KW11}, due to our different conventions.
},
\begin{equation}
 y = \frac{\r \, \cos \chi}{6} \ , \qquad w = 9 \, x_{11} + \frac{\xi}{6} \ , \qquad \tilde{\psi} = \xi
\end{equation}
and
\begin{equation}
 q =\frac{1}{36} + \frac{3}{2} \, \r^2 \, \sin^2 \chi \ .
\end{equation}
One can also check that with the above definitions the constraints \eqref{DC1}-\eqref{DC4}, proving that the solution of appendix \ref{UpliftNATDKW} fits into the class of solutions found in \cite{Gauntlett:2004zh}, are satisfied.


\section{The Abelian T-dual of the Klebanov-Witten solution}\label{appendixa}

\begin{figure}[ht]
	\begin{center}
		\includegraphics[width=10cm]{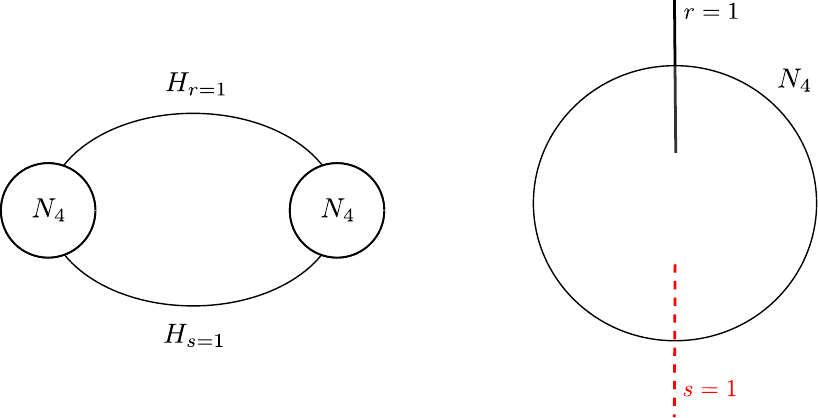}
	\end{center}
	\caption{Circular quiver associated to the Abelian T-dual solution and corresponding brane set-up. There are $N_{4}$ D4-branes stretched between the NS5 and NS5' branes. NS5 and NS5'-branes are represented by transversal black and red dashed lines, respectively.}
	\label{fig:ATDbranes1}	
\end{figure}

The Abelian T-dual, Type IIA description, of the Klebanov-Witten theory is particularly useful for the study of certain properties of this theory \cite{Dasgupta:1998su,Uranga:1998vf}. One interesting aspect is that the field theory can directly be read from the D4, NS5, NS5' brane set-up associated to this solution. We have depicted both the brane set-up and the associated quiver in Figure \ref{fig:ATDbranes1}.

In this Appendix we discuss some aspects of this description that are relevant for the understanding of the CFT interpretation of the non-Abelian T-dual solution, the main objective of this work.



\subsection{Background}

The paper  \cite{Dasgupta:1998su} considered an
Abelian T-duality transformation along the Hopf-fiber direction of the $T^{1,1}$. This dualization gives rise to a well-defined string theory background. It is however a typical example of {\it Supersymmetry without supersymmetry} \cite{Duff:1997qz}, being the low energy supergravity background non-supersymmetric.  Since our ultimate goal in this section will be to compare with the non-Abelian T-dual solution, which is only guaranteed to be a well-defined string theory background at low energies, 
we will instead dualize along the $\phi_2$ azimuthal direction of the $T^{1,1}$. This  preserves the $\mathcal{N}=1$ supersymmetry of the Klebanov-Witten solution, and can be matched directly with the non-Abelian T-dual solution in the large $\rho$ limit. 

We start by rewriting the Klebanov-Witten metric in terms of the T-duality preferred frame, in which $\phi_2$ does only appear in the form $d\phi_2$ and just in one vielbein,
\begin{align}
 & e^{x^\mu}= \frac{r}{L}dx^\mu \, , \qquad e^r= \frac{L}{r}dr \, , \qquad
 e^1= L\,\l_1\,d\theta_1 \, , \qquad e^2=L \,\l_1\,\sin\theta_1 \,d\phi_1 \, ,\nonumber
 \\[5pt]
 & e^{\hat{1}}= L\,\l_2\,d\theta_2 \, , \qquad e^{\hat 2}= L\,\l\,\l_2\,\frac{\sin\theta_2}{\sqrt{P(\theta_2)}} \, \big( d\psi + \cos\theta_1\,d\phi_1 \big) \, , \nonumber
 \\[5pt]
 & e^3= e^{C} \, \big( d\phi_2 + \tilde A_1 \big) \ ,
\end{align}
where $e^{2C}=L^2P(\theta_2)$ with $P(\theta_2)=\l^2\cos^2\theta_2 + \l_2^2 \sin^2\theta_2$, and we have introduced the connection
\begin{equation*}
 \tilde A_1= \frac{\lambda^2 \,\cos\theta_2}{P(\theta_2)} \, \big( d\psi + \cos\theta_1\,d\phi_1 \big) \;.
\end{equation*}
The Klebanov-Witten metric thus reads
\begin{align}
 ds^2 &=ds^2_{AdS_5}+ L^2 \, \Bigg[ \l_1^2\,d\Omega_2^2(\theta_1,\phi_1) + \l_2^2 \Big( d\theta_2^2 + \frac{\l^2\,\sin^2\theta_2}{P(\theta_2)} \, \big( d\psi+\cos\theta_1d\phi_1 \big)^2 \Big)  \nonumber\\
 &  \qquad\qquad + P(\theta_2) \Big( d\phi_2 + \frac{\l^2\,\cos\theta_2}{P(\theta_2)} \, \big( d\psi+\cos\theta_1d\phi_1 \big) \Big)^2 \Bigg] \;.
\end{align}

A $U(1)$ T-duality performed on the $\phi_2$ direction trades the vielbein $e^3$ for $\hat e= \a' e^{-C}d\phi_2$, and generates a NS-NS 2-form $B_2=\a'\tilde A_1\wedge d\phi_2$. The NS-NS sector for the dual solution is then given by\footnote{We rescale $\phi_2\rightarrow \frac{L^2}{\a'}\phi_ 2$, so that the metric of the internal space scales with $L^2$. We also use that $\lambda_2=\lambda_1$ for later comparison with the NATD solution.}:
\begin{align}\label{eq:ATD_NSNS}
 d s_{ATD}^2 &=ds^2_{AdS_5}+ L^2 \lambda_1^2 \Bigg[ d\Omega_2^2(\theta_1,\phi_1) +  \Big( d\theta_2^2 + \frac{\l^2\,\sin^2\theta_2}{P(\theta_2)} \, \big( d\psi+\cos\theta_1d\phi_1 \big)^2 \Big) + \frac{d\phi_2^2 }{\lambda_1^2 P(\theta_2)} \Bigg] \, , \nonumber
 \\[5pt]
 B_2^{ATD} &= -\frac{L^2\,\lambda^2 \,\cos\theta_2}{P(\theta_2)} \, \big( d\phi_2\wedge d\psi + \cos\theta_1\, d\phi_2\wedge d\phi_1 \big) \, , \nonumber
 \\[5pt]
 e^{-2\Phi_{ATD}} &= \frac{L^2}{g_s^2 \a'}\,P(\theta_2) \, .
\end{align}
We can see in the metric the geometrical realization of the $U(1)$ R-symmetry in the $\psi$ direction. 
We can also see that it agrees with the asymptotic form of the metric of the non-Abelian T-dual solution, given by the first equation in (\ref{metricasympt}), under the replacements
\begin{equation}
\label{replacements}
\chi\rightarrow \theta_2\, , \qquad \xi\rightarrow \psi\, , \qquad \rho\rightarrow \phi_2\, .
\end{equation}
The $B_2$ fields do also agree, once a gauge transformation of parameter 
\begin{equation}
\Lambda=-L^2 \cos{\theta_2}\, \phi_2 \Bigl(d\psi+\frac{\lambda^2\cos{\theta_1}}{P(\theta_2)}d\phi_1\Bigr)
\end{equation}
is performed, giving rise to 
\begin{eqnarray}
B_2 & = & -L^2\phi_2 \Bigg[ d\Omega_2(\theta_2,\psi)+ \frac{\lambda^2\cos{\theta_2}}{P(\theta_2)} \, d\Omega_2(\theta_1,\phi_1) - \lambda^2 \, \cos\theta_1 \, \partial_{\theta_2}
\Bigg(  \frac{\cos{\theta_2}}{P(\theta_2)} \Bigg) \, d\theta_2\wedge d\phi_1  \Bigg] \nonumber
\\[5pt]
& & \qquad \qquad \qquad  +\frac{L^2 \sin{\theta_2}}{2\,P(\theta_2)} \, \big( \l^2-\l_1^2 \big) \, \sin 2\theta_2 \, d\psi\wedge d\phi_2 \, .
\end{eqnarray}
We will use this expression for the $B_2$-field in the remaining of this section. As in \cite{Lozano:2016kum,Lozano:2016wrs}, the two dilatons satisfy $e^{-2\Phi_{NATD}} \approx \r^2\, e^{-2\Phi_{ATD}}$ for large $\rho$ (after re-absorbing the scaling factors in $\rho \to \frac{\a'}{L^2}\r$). As explained in  \cite{Lozano:2016kum,Lozano:2016wrs}, this relation has its origin in the different measures in the partition functions of the non-Abelian and Abelian T-dual sigma models.

Finally, the RR fields are:
\begin{equation}
 \begin{aligned}
  & F_4 = \frac{4 \, L^4 \, \l \, \l_1^4}{g_s \, \alpha'^{1/2}} \, \sin\theta_1 \sin\theta_2 \ d\theta_1 \wedge d\phi_1\wedge d\theta_2 \wedge d\psi \ ,
  \\[5pt]
  & F_6 = \frac{4 \, L}{g_s \, \alpha'^{1/2}} \, \textrm{Vol}_{AdS_5} \wedge d\phi_2 \, .
 \end{aligned}
\end{equation}
One can check that, as in \cite{Lozano:2016wrs}, for large $\rho$ the fluxes polyforms satisfy 
\begin{equation}
e^{\Phi_{NATD}}F_{NATD}\approx e^{\Phi_{ATD}}F_{ATD}\, .
\end{equation}

The previous relations show that the non-Abelian T-dual solution reduces in the $\rho\rightarrow\infty$ limit to the Abelian T-dual one. This connection between non-Abelian and Abelian T-duals was discussed previously in examples where the dualization took place on a round $S^3$ \cite{Lozano:2016kum,Lozano:2016wrs}. Our results show that it extends more generally.  It is worth stressing however that in this case the relation is more subtle globally. 
Indeed, the relations in eq.\! (\ref{replacements}) identify $\xi\in [0,2\pi]$ with $\psi\in [0,4\pi]$. The reason for this apparent mismatch is that the dualization on $\phi_2$ generates a bolt singularity in the metric, and this must be cured by setting $\psi\in [0,2\pi]$, such that the bolt singularity reduces to the coordinate singularity of $\mathbb{R}^2$ written in polar coordinates. Once this is taken into account the ranges of both coordinates also agree. As encountered in  \cite{Itsios:2013wd}, the dualization has enforced a $\mathbb{Z}_2$ quotient on $\psi$. Our Abelian T-dual is thus describing the Klebanov-Witten theory modded by  $\mathbb{Z}_2$. This is consistent with the brane set-up that is implied by the quantized charges of the background, as we now show.


\subsection{Quantized charges and brane set-up}

The background fluxes of the Abelian T-dual solution support $D4$ and $NS5$-brane charges. The Page charge for the $D4$ branes is given by:
\begin{equation}
 Q_{D4} = \frac{1}{2 \, \kappa^2_{10} \, T_{D4}} \int_{M_4} F_4 = \frac{2}{27} \frac{L^4}{\pi \, g_s^2 \, \alpha'^2} = N_4 \ .
\end{equation}
Imposing the quantization of this charge we find that the radius $L$ is related to the number of $D4$ branes through the formula:
\begin{equation}
\label{LtoN4}
 L^4 = \frac{27}{2} \, \pi \, g_s^2 \, \alpha'^2 \, N_4 \ .
\end{equation}
We find a factor of 2 of difference with respect to the original background. This is due to the 
change in the periodicity of the $\psi$ direction from $[0,4\pi]$ to $[0,2\pi]$.

In turn, the charge of $NS5$ branes is calculated from:
\begin{equation}
 Q_{NS5} =  \frac{1}{ 4 \, \pi^2 \, \alpha'} \int_{M_3} H_3\, . 
\end{equation}
As in section \ref{quantized-charges}, we can define two 3-cycles:  $\Sigma_3=[\phi_2,\theta_2,\psi]$  and 
$\Sigma_3'=[\phi_2,\theta_1,\phi_1]_{\theta_2=0}$. Taking $M_3$ to be any of these cycles we find that there are two units of NS5, or NS5', charge. This is consistent with a brane picture of two alternating NS5, NS5' branes, transverse to  either of the two 2-cycles ${\tilde S}^2(\theta_2,\psi)$, $S^2(\theta_1,\phi_1)$, located along the compact $\phi_2$-direction. This is the brane set-up discussed in \cite{Uranga:1998vf}, describing the Klebanov-Witten theory modded by $\mathbb{Z}_2$ in Type IIA. The general $\mathbb{Z}_k$ case is depicted in Figure \ref{fig:ATDbranes} of Appendix \ref{ATDZk}.
Note that, as discussed in \cite{Uranga:1998vf}, the positions of the branes in the $\phi_2$-circle are not specified by the geometry, so generically we can only think that they define four intervals in the $\phi_2$-circle \footnote{In \cite{Uranga:1998vf} it was argued that the different orderings correspond to different phases in the K\"ahler moduli space of the orbifold singularity. This is interpreted in the field theory side in terms of Seiberg duality \cite{Elitzur:1997fh}, so the corresponding theories should flow to the same CFT in the infrared.}.
The same number of D4-branes are stretched between each pair of NS5, NS5' branes since even if large gauge transformations are required as we pass the value $\phi_2=\pi L^2/\a'$, the D4-brane charge does not change in the absence of $F_2$-flux. 

Coming back to Section \ref{centralcharge}, the relation found there between the central charges of the non-Abelian and Abelian T-dual solutions helps us understand now the connection between $\rho$ and $\phi_2$ globally. The computation in that section showed that the central charges agree when $\rho\in [n\pi\frac{L^2}{\alpha'}, (n+1)\pi\frac{L^2}{\alpha'}]$ and $n$ is sent to infinity. This is consistent with the $\rho\rightarrow\infty$ limit that must be taken at the level of the solutions. Furthermore, it clarifies why globally the $\rho$ direction is identified, through the replacements in (\ref{replacements}), with $\phi_2\in [0,2\pi \frac{L^2}{\alpha'}]$. This is just implied by the $\mathbb{Z}_2$ quotient enforced by the Abelian T-duality transformation.

\section{Some field theory elaborations}\label{appendixb}
In this appendix we discuss some aspects 
of the field theory analysis presented in Section \ref{quiverNATD}. We start with the calculation of the beta functions and anomalies for the Klebanov-Witten CFT.

\subsection{A summary of the Klebanov-Witten CFT}

The field content of the Klebanov-Witten theory consists on a $SU(N)\times SU(N)$ gauge group with bifundamental matter fields 
$A_1 , \, A_2$ and $B_1 , \, B_2$, transforming in the $(N , \bar{N})$ and $(\bar{N}, N)$ representations of $SU(N)$, respectively. This theory is represented by the quiver depicted in Figure \ref{Anomaly}.
\begin{figure}[h]
	\centering
	\includegraphics[scale=0.35]{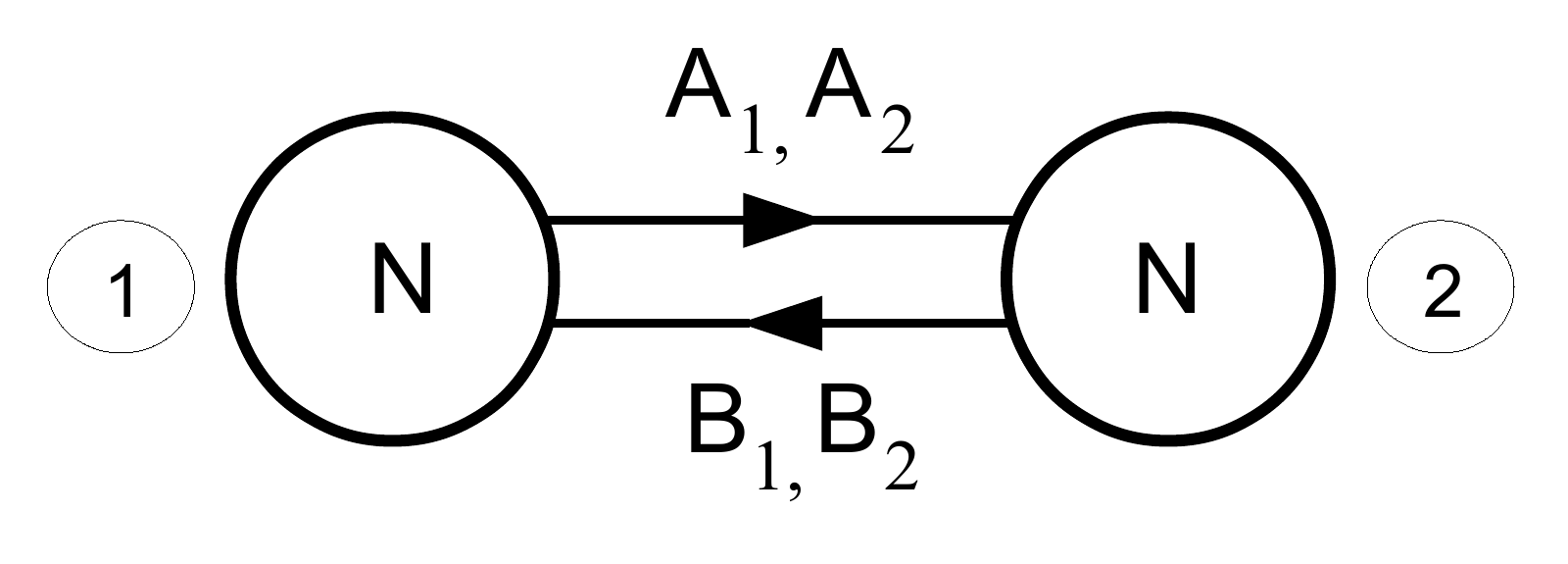}
	\caption{The KW quiver.}
\label{Anomaly}
\end{figure}
The anomalous dimensions of the matter fields are,
\begin{equation}
 \g_{A_i} = \g_{B_i} = - \frac{1}{2} \ ,
\end{equation}
and thus the physical dimensions and the R-charges are given by,
\begin{equation}
 \begin{aligned}
  & \textrm{dim}(A_i) =  \textrm{dim}(B_i) = 1 - \frac{1}{4} = \frac{3}{4} \ ,
 \\[10pt]
  & R_{[A]} = R_{[B]} = \frac{1}{2} \ ,\qquad R_{\Psi_A} = R_{\Psi_B} = - \frac{1}{2} \ .
 \end{aligned}
\end{equation}
Substituting in eq. (\ref{beta-function}) we see that the $\beta$-functions for the couplings $g_1$ and $g_2$ vanish:
\begin{equation}
 \begin{aligned}
  & \beta_{i} \sim 3 \, N - 2 \, N \, \Big[   1 - \Big(  - \frac{1}{2}  \Big)  \Big] = 0 \ , \qquad i = 1, 2 \ .
 \end{aligned}
\end{equation}
We can also check the vanishing of the anomaly,
\begin{equation}
 \D \theta_i = 2 \, N + 2 \, (2 \, N) \, \Big(  - \frac{1}{2}  \Big) = 0 \ ,
\end{equation}
where we took into account that the R-charge of the gaugino is $1$ while that of the two Weyl fermions is $- 1/2$.

We hope that this has prepared the reader unfamiliar with these formalities to understand the material in our Section \ref{quiverNATD}.

\subsection{Central Charge of the ${\cal N}=2$ UV CFT} \label{centralchargeN=2}

In this Appendix we compute the central charge of the $\mathcal{N}=2$ quiver associated to the non-Abelian T-dual of $AdS_5\times S^5/\mathbb{Z}_2$, using $a$-maximization. 
We obtain that the central charge is maximized for $\epsilon=\frac13$, as for the equal rank quivers considered in  \cite{Bah:2013aha}. Furthermore, we show that the result of this calculation leads, consistently, to the holographic central charge given by eq.\! (\ref{papa}).
 
We consider the $\mathbb{Z}_2$-reflection of the quiver of Figure \ref{Sfetsos-Thompson-quiver} and take $\sigma_i=+1$ for all hypermultiplets, including the ones associated with the flavor groups.
 We  then find for the trace anomalies ($N\equiv N_6$):
\begin{align}\label{TrRH} 
  \Tr\, R_\epsilon(H)
   &= 2\sum_{j=1}^{n-1} \Tr\, R_\epsilon(H_j)  + \sum_{i=1}^2 \Tr\, R_\epsilon(H_{f_i}) \nonumber
   \\[5pt]
   &=2N^2 \sum_{j=1}^{n-1} \, j \, \big( j + 1 \big) \, \big( \e \, \s_j - 1 \big)  + n \, N^2 \sum_{i=1}^2 \big( \e \, \s_{f_i} - 1 \big)  \nonumber
   \\[5pt]
   &= N^2 \left[ \frac23 \,(n^3-n)\big( \e - 1 \big) +  2n (\e-1) \right]  \nonumber
   \\[7pt]   
   & = N^2 \left[ \frac23 n^3 + \frac43 n \right]\big( \e - 1 \big) \approx \frac{2}{3} \, n^3 \, N^2 \, \big( \epsilon - 1 \big) + \cO(n) \, , 
\end{align}
as well as
\begin{align}
  \Tr\, R^3_\epsilon(H)
   &= 2\sum_{j=1}^{n-1} \Tr\, R^3_\epsilon(H_j) + \sum_{i=1}^2 \Tr\, R^3_\epsilon(H_{f_i}) \nonumber
    \\[5pt]
   &= 2 \, N^2 \sum_{j=1}^{n-1} j \, \big( j + 1 \big) \, \frac{ \big( \e - 1 \big)^3}{4}  + n \, N^2 \sum_{i=1}^2 \frac{ \big( \e \, \s_{f_i} - 1 \big)^3}{4} \nonumber
     \\[5pt]
   & = \frac{N^2}{4} \Bigg[ \frac23 \,  (n^3-n) \big( \e - 1 \big)^3  
      + 2n\big(\e-1 \big)^3 \Bigg] \nonumber
     \\[5pt]
   & = N^2 \left[ \frac16 n^3 + \frac13 n \right]\big( \e - 1 \big)^3\approx \frac{1}{6} \, n^3 \, N^2 \big( \epsilon - 1 \big)^3 + \cO(n) \, .
\end{align}

For the $\cN=2$ vector multiplets ($\cN=1$ vector $+$ chiral adjoint) the non-anomalous R-charge $R_\e=R_0+\epsilon \, \cF/2$ is obtained from the R-charge for the gaugino, $R_0(\l)=1$, plus the non-anomalous flavor charge of the fermion in the chiral adjoint $\cF(\psi_j)=(-1) \big( \s_{j-1} + \s_j \big)$, being $R_0(\psi_j)=0$. We thus have:
\begin{align}\label{N_2_vector_j}
\Tr\,R_\epsilon(V_j) &=  \big( N_j^2 - 1 \big) \, \Big( 1-\frac{1}{2} \, \e \, \big( \s_{j-1} + \s_j \big) \Big) \,, \nonumber
\\[5pt]
\Tr\,R^3_\epsilon(V_j) &= \big( N_j^2 - 1 \big) \, \Big( 1-\frac{1}{8} \, \e^3 \, \big( \s_{j-1} + \s_j \big)^3 \Big)\,.
\end{align}
These are summed up easily for all $\s_j=+1$:
\begin{align}\label{TrRV}
\Tr\, R_\epsilon(V) &=2 \sum_{j=1}^{n-1} \Tr\, R_\epsilon(V_j) + \Tr\, R_\epsilon(V_n)= \left[2 \sum_{j=1}^{n-1} \big( j^2N^2 - 1 \big) + \big( n^2N^2 - 1 \big) \right] \Big( 1- \e\Big) \nonumber
\\[5pt]
&= \Big[ \frac13\left( 2n^3+n \right)N^2 -(2n-1)  \Big]\big(1-\e\big) \approx \frac{2}{3} \, n^3 \, N^2 \big( 1 - \e \big) + \cO(n) \, . 
\end{align} 
The cubic term follows most readily:
\begin{equation}
 \Tr\, R^3_\epsilon(V) = \Big[ \frac13\left( 2n^3+n \right)N^2 -(2n-1)  \Big]\big(1-\e^3\big) \approx \frac{2}{3} \, n^3 \, N^2 \, \big( 1 - \e^3 \big) + \cO(n)\;.
\end{equation}
We thus see that both linear contributions \eqref{TrRH} and \eqref{TrRV} from the hypermultiplets and vector multiplets cancel at leading order, so that
\[ \Tr\,R_\e \equiv \Tr\,R_\e(H) + \Tr\,R_\e(V) \approx \cO(n) \;.\]
Now both $a(\e)$ and $c(\e)$ charges can be computed exactly,
\begin{align}
a(\e) &= \frac{3}{64}\,(1-\e) \Bigg\{ \Big[ 3n^3(1+\e)^2  + 2(1+3\e)n \Big]N_6^2  - 2(2n-1)\Big(2+3\e(1+\e)\Big) \Bigg\}, \nonumber \\[5pt]
c(\e) &= \frac{1}{64}\,(1-\e) \Bigg\{ \Big[ 9n^3(1+\e)^2 + 2(5+9\e)n \Big]N_6^2 - 2(2n-1)\Big(4+9\e(1+\e)\Big) \Bigg\},
\end{align}
and $a(\e)$ is maximized for $\e=1/3$, yielding the superconformal charges:
\begin{align}\label{eq:exact_N2_ac}
a_{\cN=2} \equiv a(\e=1/3) &= \frac{1}{24}\left\{  (4n^3 + 3n)N_6^2 -10n + 5 \right\} , \nonumber \\[5pt]
c_{\cN=2} \equiv c(\e=1/3) &=\frac{1}{6}\left\{  (n^3 + n)N_6^2 -2n +1\right\}.
\end{align}  
In the long quiver approximation, we recover the holographic result
\begin{equation}
  c_{\cN=2} \approx a_{\cN=2} \approx \frac{1}{6} \, n^3 \, N_6^2  + \cO \big( n \big) \,,
\end{equation} 
as expected. It is noteworthy that $\e=1/3$ is the value of $\epsilon$ predicted in \cite{Bah:2013aha} for $\mathcal{N}=2$ quivers with nodes of the same rank. 

\subsection{Central charge of the Klebanov-Witten theory modded by $\mathbb{Z}_k$} 
\label{ATDZk}

In this Appendix we include, for completeness, the field theory calculation of the central charge of the Klebanov-Witten theory, using $a$-maximization. We will center in the more general case in which the theory is modded by $\mathbb{Z}_k$. The computation of the field theoretical central charge in this example is very illustrative of the $a$-maximization technique used throughout the paper. 

\begin{figure}[ht]
	\begin{center}
		\includegraphics[width=16cm]{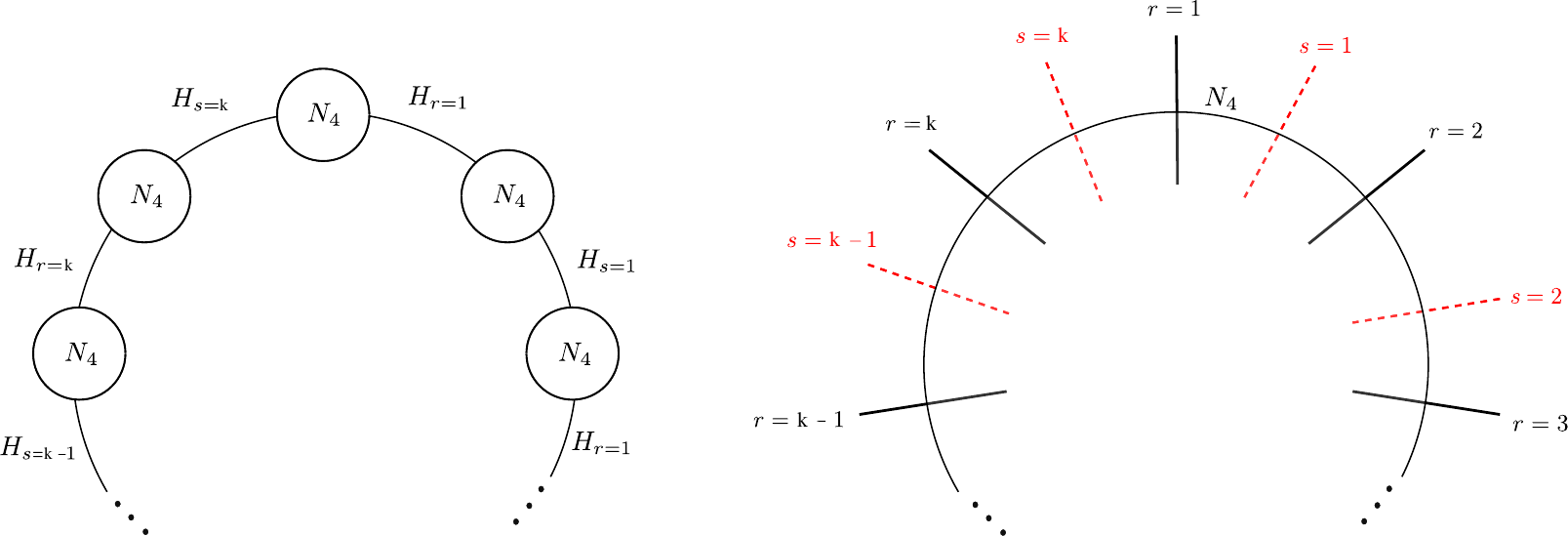}
	\end{center}
	\caption{Circular quiver associated to the KW theory modded by $\mathbb{Z}_k$, and corresponding brane set-up. There are $N_{4}$ D4-branes stretched between $p=k$ NS5-branes, labeled by $r=1,\ldots,k$ (as for the corresponding hypermultiplets) and $q=k$ NS5'-branes labeled by  $s=1,\ldots,k$. NS5 and NS5'-branes are represented by transversal black and red dashed lines, respectively.}
	\label{fig:ATDbranes}	
\end{figure}

In this case we have, in the Type IIA description, 
$p=k$ NS5-branes and $q=k$ NS5'-branes, and $\ell=p+q=2k$ hypermultiplets connecting $\ell$ $\cN=1$ vector multiplets \cite{Uranga:1998vf}.
The first and the last nodes are made to coincide, as depicted in \autoref{fig:ATDbranes}.
We closely follow the field-theoretical computation of the central charge for the linear quiver proposed in Section \ref{sec:cc_compu}. We just need to take $N_a=N_b=N_{4}$ for the bifundamentals in \eqref{eq:hypers} for all the $\ell=2k$ nodes. This yields the linear contribution for the hypermultiplets:
\begin{align*}
\Tr\, R_\epsilon(H) &\equiv \sum_{j=1}^{\ell} \Tr\, R_\epsilon(H_j) = \sum_{r=1}^{k} \Tr\, R_\epsilon(H_r) + \sum_{s=1}^{k} \Tr\, R_\epsilon(H_s)
\\[5pt]
& = \ell\,N_{4}^2 \big( z \, \epsilon-1 \big) \stackrel{z=0}{=} -2 \, k \, N_{4}^2 \;,
\end{align*}
where we have used $z=(p-q)/\ell$. Similarly, the cubic contribution is given by
\begin{equation*}
\Tr\, R^3_\epsilon(H) = \frac{\ell}{4}\,N_{4}^2\Big( z \, \big( \epsilon^3 + 3 \, \epsilon \big)-3 \, \epsilon^2-1 \Big) \Big|_{z=0} = - \frac{k}{2} \, N_{4}^2 \, \big( 3 \, \epsilon^2 + 1 \big) \;.
\end{equation*}
Contributions from $\cN=1$ vector multiplets are computed straightforwardly to be:
\begin{equation*}
\Tr\, R_\epsilon(V) = \Tr\, R^3_\epsilon(V)=\ell \, \big( N_{4}^2-1 \big) = 2 \, k \, \big( N_{4}^2-1 \big) \, .
\end{equation*}
We can now use \eqref{eq:ac_charges} to get
\begin{equation}
  c(\epsilon) = \frac{\ell}{128} \, \Big( 27 \, N_{4}^2 \, \big( 1 - \epsilon^2 \big) - 16 \Big) \, ,
\end{equation}
which is maximized for $\epsilon=0$, yielding the fixed point central charge (for large $N_{4}$):
\begin{equation}
\label{eq:cc_ATD_FT}
c \approx \frac{27}{64} \, k \, N_{4}^2 \, ,
\end{equation}
which coincides, as expected, with the holographic value (given by eq.\! (\ref{abeliancc}) for $k=1$). This expression is valid for any $k\geq1$, i.e. no large $\ell=2k$ limit has been assumed. 

Note that in the absence of flavor groups it is not possible to define $\sigma_0,\,\sigma_\ell$, and neither $\k=\sigma_0+\sigma_\ell$, as we have done for the linear quivers discussed in section \ref{subsec_linear_quivers}. Still, the result in (\ref{eq:cc_ATD_FT}) agrees with the central charge of a Bah-Bobev type of linear quiver (see eq. (3.20) in \cite{Bah:2013aha}) for $\kappa=0$ and large $\ell$. Indeed, even if there is no clear definition for $\k$ in this case, the uplift of the circular brane set-up is interpreted as M5-branes wrapping a torus ($\k=0$) with minimal punctures, as the gauging of the end flavor groups of the linear quiver corresponds in M-theory to gluing the two left-over maximal punctures, closing up the Riemann surface.



\begin{thebibliography}{99}


\bibitem{delaOssa:1992vci} 
  X.~C.~de la Ossa and F.~Quevedo,
  {\it Duality symmetries from nonAbelian isometries in string theory},
  Nucl.\ Phys.\ B {\bf 403}, 377 (1993),
  \href{http://arxiv.org/abs/hep-th/9210021}{{\tt hep-th/9210021}}.
  
\bibitem{Alvarez:1993qi} 
  E.~Alvarez, L.~Alvarez-Gaume, J.~L.~F.~Barbon and Y.~Lozano,
  {\it Some global aspects of duality in string theory},
  Nucl.\ Phys.\ B {\bf 415}, 71 (1994),
  \href{http://arxiv.org/abs/hep-th/9309039}{{\tt hep-th/9309039}}.
  E.~Alvarez, L.~Alvarez-Gaume and Y.~Lozano,
  {\it On nonAbelian duality},
  Nucl.\ Phys.\ B {\bf 424}, 155 (1994),
  \href{http://arxiv.org/abs/hep-th/9403155}{{\tt hep-th/9403155}}.
  E.~Alvarez, L.~Alvarez-Gaume and Y.~Lozano,
  {\it A Canonical approach to duality transformations},
  Phys.\ Lett.\ B {\bf 336}, 183 (1994),
  \href{http://arxiv.org/abs/hep-th/9406206}{{\tt hep-th/9406206}}.
  S.~Elitzur, A.~Giveon, E.~Rabinovici, A.~Schwimmer and G.~Veneziano,
  {\it Remarks on nonAbelian duality},
  Nucl.\ Phys.\ B {\bf 435}, 147 (1995),
  \href{http://arxiv.org/abs/hep-th/9409011}{{\tt hep-th/9409011}}.
  C.~Klimcik and P.~Severa,
  {\it Dual nonAbelian duality and the Drinfeld double},
  Phys.\ Lett.\ B {\bf 351}, 455 (1995),
  \href{http://arxiv.org/abs/hep-th/9502122}{{\tt hep-th/9502122}}.
  Y.~Lozano,
  {\it NonAbelian duality and canonical transformations},
  Phys.\ Lett.\ B {\bf 355}, 165 (1995),
  \href{http://arxiv.org/abs/hep-th/9503045}{{\tt hep-th/9503045}}.
  K.~Sfetsos, 
  {\it NonAbelian duality, parafermions and supersymmetry}, Phys.\ Rev. \ D {54}, 1682 (1996), 
   \href{http://arxiv.org/abs/hep-th/9602179}{{\tt hep-th/9602179}}.

  
  







\bibitem{Sfetsos:2010uq}
  K.~Sfetsos and D.~C.~Thompson,
  {\it On non-abelian T-dual geometries with Ramond fluxes},
  Nucl.\ Phys.\ B {\bf 846} (2011) 21,
  \href{http://arxiv.org/abs/arXiv:1012.1320}{{\tt arXiv:1012.1320}}.




\bibitem{Maldacena:1997re}
  J.~M.~Maldacena,
  {\it The Large N limit of superconformal field theories and supergravity},
  Int.\ J.\ Theor.\ Phys.\  {\bf 38} (1999) 1113,
   [Adv.\ Theor.\ Math.\ Phys.\  {\bf 2} (1998) 231],
  \href{http://arxiv.org/abs/hep-th/9711200}{{\tt hep-th/9711200}}.

\bibitem{Lozano:2012au}
  Y.~Lozano, E.~O Colgain, D.~Rodriguez-Gomez and K.~Sfetsos,
  {\it Supersymmetric $AdS_6$ via T Duality},
  Phys.\ Rev.\ Lett.\  {\bf 110} (2013) no.23,  231601,
  \href{http://arxiv.org/abs/arXiv:1212.1043}{{\tt arXiv:1212.1043}}.

\bibitem{Itsios:2013wd}
  G.~Itsios, C.~Nunez, K.~Sfetsos and D.~C.~Thompson,
  {\it Non-Abelian T-duality and the AdS/CFT correspondence:new N=1 backgrounds},
  Nucl.\ Phys.\ B {\bf 873} (2013) 1,
  \href{http://arxiv.org/abs/arXiv:1301.6755}{{\tt arXiv:1301.6755}}.
  G.~Itsios, C.~Nunez, K.~Sfetsos and D.~C.~Thompson,
  {\it On Non-Abelian T-Duality and new N=1 backgrounds},
  Phys.\ Lett.\ B {\bf 721} (2013) 342,
  \href{http://arxiv.org/abs/arXiv:1212.4840}{{\tt arXiv:1212.4840}}.

  


 \bibitem{variosa2}
   Y.~Lozano, E.~O Colgain, K.~Sfetsos and D.~C.~Thompson,
  {\it Non-abelian T-duality, Ramond Fields and Coset Geometries},
  JHEP {\bf 1106} (2011) 106,
  \href{http://arxiv.org/abs/arXiv:1104.5196}{{\tt arXiv:1104.5196}}.
  G.~Itsios, Y.~Lozano, E.~O Colgain and K.~Sfetsos,
  {\it Non-Abelian T-duality and consistent truncations in type-II supergravity},
  JHEP {\bf 1208} (2012) 132,
  \href{http://arxiv.org/abs/arXiv:1205.2274}{{\tt arXiv:1205.2274}}.
  J.~Jeong, O.~Kelekci and E.~O Colgain,
  {\it An alternative IIB embedding of F(4) gauged supergravity},
  JHEP {\bf 1305}, 079 (2013),
  \href{http://arxiv.org/abs/arXiv:1302.2105}{{\tt arXiv:1302.2105}}.
  A.~Barranco, J.~Gaillard, N.~T.~Macpherson, C.~Nunez and D.~C.~Thompson,
  {\it G-structures and Flavouring non-Abelian T-duality},
  JHEP {\bf 1308}, 018 (2013),
  \href{http://arxiv.org/abs/arXiv:1305.7229}{{\tt arXiv:1305.7229}}.
  N.~T.~Macpherson,
  {\it Non-Abelian T-duality, $G_2$-structure rotation and holographic duals of $N=1$ Chern-Simons theories},
  JHEP {\bf 1311}, 137 (2013),
  \href{http://arxiv.org/abs/arXiv:1310.1609}{{\tt arXiv:1310.1609}}.
  
  \bibitem{Lozano:2013oma} 
  Y.~Lozano, E.~O.~Colgain and D.~Rodriguez-Gomez,
  {\it Hints of 5d Fixed Point Theories from Non-Abelian T-duality},
  JHEP {\bf 1405}, 009 (2014),
  \href{http://arxiv.org/abs/arXiv:1311.4842}{{\tt arXiv:1311.4842}}.

  \bibitem{bbranch}
  J.~Gaillard, N.~T.~Macpherson, C.~Nunez and D.~C.~Thompson,
  {\it Dualising the Baryonic Branch: Dynamic SU(2) and confining backgrounds in IIA},
  Nucl.\ Phys.\ B {\bf 884}, 696 (2014),
  \href{http://arxiv.org/abs/arXiv:1312.4945}{{\tt arXiv:1312.4945}}.

\bibitem{variosb1} 
  D.~Elander, A.~F.~Faedo, C.~Hoyos, D.~Mateos and M.~Piai,
  {\it Multiscale confining dynamics from holographic RG flows},
  JHEP {\bf 1405}, 003 (2014),
  \href{http://arxiv.org/abs/arXiv:1312.7160}{{\tt arXiv:1312.7160}}.
  S.~Zacarias,
  {\it Semiclassical strings and Non-Abelian T-duality},
  Phys.\ Lett.\ B {\bf 737}, 90 (2014),
  \href{http://arxiv.org/abs/arXiv:1401.7618}{{\tt arXiv:1401.7618}}.
  E.~Caceres, N.~T.~Macpherson and C.~Nunez,
  {\it New Type IIB Backgrounds and Aspects of Their Field Theory Duals},
  JHEP {\bf 1408}, 107 (2014),
  \href{http://arxiv.org/abs/arXiv:1402.3294}{{\tt arXiv:1402.3294}}.
  P.~M.~Pradhan,
  {\it Oscillating Strings and Non-Abelian T-dual Klebanov-Witten Background},
  Phys.\ Rev.\ D {\bf 90}, no. 4, 046003 (2014),
  \href{http://arxiv.org/abs/arXiv:1406.2152}{{\tt arXiv:1406.2152}}.
  
\bibitem{Lozano:2014ata}
  Y.~Lozano and N.~T.~Macpherson,
  {\it A new AdS$_{4}$/CFT$_{3}$ dual with extended SUSY and a spectral flow},
  JHEP {\bf 1411} (2014) 115,
  \href{http://arxiv.org/abs/arXiv:1408.0912}{{\tt arXiv:1408.0912}}.
  
  
  \bibitem{variosa3}
  K.~Sfetsos and D.~C.~Thompson,
  {\it New ${\cal N} = 1$ supersymmetric $AdS_5$ backgrounds in Type IIA supergravity},
  JHEP {\bf 1411}, 006 (2014),
  \href{http://arxiv.org/abs/arXiv:1408.6545}{{\tt arXiv:1408.6545}}.
  O.~Kelekci, Y.~Lozano, N.~T.~Macpherson and E.~O.~Colgain,
  {\it Supersymmetry and non-Abelian T-duality in type II supergravity},
  Class.\ Quant.\ Grav.\  {\bf 32} (2015) no.3,  035014,
  \href{http://arxiv.org/abs/arXiv:1409.7406}{{\tt arXiv:1409.7406}}.
  
  \bibitem{Macpherson:2014eza} 
  N.~T.~Macpherson, C.~Nunez, L.~A.~Pando Zayas, V.~G.~J.~Rodgers and C.~A.~Whiting,
  {\it Type IIB supergravity solutions with AdS$_{5}$ from Abelian and non-Abelian T dualities},
  JHEP {\bf 1502}, 040 (2015),
  \href{http://arxiv.org/abs/arXiv:1410.2650}{{\tt arXiv:1410.2650}}.

\bibitem{variosa4}
   K.~S.~Kooner and S.~Zacarias,
  {\it Non-Abelian T-Dualizing the Resolved Conifold with Regular and Fractional D3-Branes},
  JHEP {\bf 1508}, 143 (2015),
  \href{http://arxiv.org/abs/arXiv:1411.7433}{{\tt arXiv:1411.7433}}.
  T.~R.~Araujo and H.~Nastase,
  {\it $\mathcal{N}=1$ SUSY backgrounds with an AdS factor from non-Abelian T duality},
  Phys.\ Rev.\ D {\bf 91}, no. 12, 126015 (2015),
  \href{http://arxiv.org/abs/arXiv:1503.00553}{{\tt arXiv:1503.00553}}.
  
\bibitem{Bea:2015fja}
  Y.~Bea, J.~D.~Edelstein, G.~Itsios, K.~S.~Kooner, C.~Nunez, D.~Schofield and J.~A.~Sierra-Garcia,
  {\it Compactifications of the Klebanov-Witten CFT and new AdS$_{3}$ backgrounds},
  JHEP {\bf 1505} (2015) 062,
\href{http://arxiv.org/abs/arXiv:1503.07527}{{\tt arXiv:1503.07527}}.  
  
  
  \bibitem{Lozano:2015bra} 
  Y.~Lozano, N.~T.~Macpherson, J.~Montero and E.~O.~Colgain,
  {\it New $AdS_3 \times S^2$ T-duals with $ \mathcal{N}=\left(0,4\right) $ supersymmetry},
  JHEP {\bf 1508}, 121 (2015),
  \href{http://arxiv.org/abs/arXiv:1507.02659}{{\tt arXiv:1507.02659}}.
  
   \bibitem{Lozano:2015cra} 
  Y.~Lozano, N.~T.~Macpherson and J.~Montero,
  {\it A $ \mathcal{N}=2 $ supersymmetric AdS$_{4}$ solution in M-theory with purely magnetic flux},
  JHEP {\bf 1510} (2015) 004,
  \href{http://arxiv.org/abs/arXiv:1507.02660}{{\tt arXiv:1507.02660}}.
%

\bibitem{Araujo:2015dba} 
  T.~R.~Araujo and H.~Nastase,
  {\it Non-Abelian T-duality for nonrelativistic holographic duals},
  JHEP {\bf 1511}, 203 (2015),
  \href{http://arxiv.org/abs/arXiv:1508.06568}{{\tt arXiv:1508.06568}}.
  
\bibitem{Macpherson:2015tka}
  N.~T.~Macpherson, C.~Nunez, D.~C.~Thompson and S.~Zacarias,
  {\it Holographic Flows in non-Abelian T-dual Geometries},
  JHEP {\bf 1511} (2015) 212,
  \href{http://arxiv.org/abs/arXiv:1509.04286}{{\tt arXiv:1509.04286}}.
  
  \bibitem{variosa5}
 L.~A.~P.~Zayas, V.~G.~J.~Rodgers and C.~A.~Whiting,
  {\it Supergravity solutions with AdS$_{4}$ from non-Abelian T-dualities},
  JHEP {\bf 1602}, 061 (2016),
  \href{http://arxiv.org/abs/arXiv:1511.05991}{{\tt arXiv:1511.05991}}.
 L.~A.~Pando Zayas, D.~Tsimpis and C.~A.~Whiting,
  {\it A Supersymmetric IIB Background with $AdS_4$ from Massive IIA},
  \href{https://arxiv.org/abs/1701.01643}{{arXiv:1701.01643}}.

 

\bibitem{Apruzzi:2014qva}
  F.~Apruzzi, M.~Fazzi, A.~Passias, D.~Rosa and A.~Tomasiello,
  {\it AdS$_{6}$ solutions of type II supergravity},
  JHEP {\bf 1411} (2014) 099,
   Erratum: [JHEP {\bf 1505} (2015) 012],
 \href{http://arxiv.org/abs/arXiv:1406.0852}{{\tt arXiv:1406.0852}}.   

\bibitem{Bah:2015nva}
  I.~Bah and V.~Stylianou,
  {\it Gravity duals of N=(0,2) SCFTs from M5-branes},
   \href{http://arxiv.org/abs/arXiv:1508.04135}{{\tt arXiv:1508.04135}}.

\bibitem{Kelekci:2016uqv}
  O.~Kelekci, Y.~Lozano, J.~Montero, E.~O~Colgain and M.~Park,
  {\it Large superconformal near-horizons from M-theory},
  Phys.\ Rev.\ D {\bf 93} (2016) no.8,  086010,
   \href{http://arxiv.org/abs/arXiv:1602.02802}{{\tt arXiv:1602.02802}}.

\bibitem{Couzens:2016iot}
  C.~Couzens,
  {\it Supersymmetric AdS$_{5}$ solutions of type IIB supergravity without D3 branes},
  JHEP {\bf 1701} (2017) 041,
   \href{http://arxiv.org/abs/arXiv:1609.05039}{{\tt arXiv:1609.05039}}.

\bibitem{3d3d}
Y.~Terashima and M.~Yamazaki, {\it SL(2,R) Chern-Simons, Liouville and Gauge Theory on Duality Walls}, JHEP {\bf 1108} (2011) 135, 
 \href{http://arxiv.org/abs/arXiv:1103.5748}{{\tt arXiv:1103.5748}};
 {\it Semiclassical Analysis of the 3d/3d Relation},  Phys. Rev. {\bf D88} (2013) 2, 026011,
  \href{http://arxiv.org/abs/arXiv:1106.3066}{{\tt arXiv:1106.3066}}.
  
\bibitem{Lozano:2016kum}
  Y.~Lozano and C.~Nunez,
  {\it Field Theory Aspects of non-Abelian T-duality and N=2 Linear Quivers},
  JHEP {\bf 1605} (2016) 107,
  \href{http://arxiv.org/abs/arXiv:1603.04440}{{\tt arXiv:1603.04440}}.

\bibitem{Lozano:2016wrs}
  Y.~Lozano, N.~T.~Macpherson, J.~Montero and C.~Nunez,
  {\it Three-dimensional N=4 Linear Quivers and non-Abelian T-duals},
  JHEP {\bf 1611} (2016) 133,
  \href{http://arxiv.org/abs/arXiv:1609.09061}{{\tt arXiv:1609.09061}}.

\bibitem{Lozano:2017ole}
  Y.~Lozano, C.~Nunez and S.~Zacarias,
  {\it BMN Vacua, Superstars and Non-Abelian T-duality},
  \href{http://arxiv.org/abs/arXiv:1703.00417}{{\tt arXiv:1703.00417}}.

\bibitem{Maldacena:2000mw} 
  J.~M.~Maldacena and C.~Nunez,
  {\it Supergravity description of field theories on curved manifolds and a no go theorem},
  Int.\ J.\ Mod.\ Phys.\ A {\bf 16}, 822 (2001),
  \href{http://arxiv.org/abs/hep-th/0007018}{{\tt hep-th/0007018}}.



\bibitem{Assel:2011xz} 
  B.~Assel, C.~Bachas, J.~Estes and J.~Gomis,
  {\it Holographic Duals of D=3 N=4 Superconformal Field Theories},
  JHEP {\bf 1108}, 087 (2011),
  \href{http://arxiv.org/abs/arXiv:1106.4253}{{\tt arXiv:1106.4253}}.
  B.~Assel, C.~Bachas, J.~Estes and J.~Gomis,
  {\it IIB Duals of D=3 N=4 Circular Quivers},
  JHEP {\bf 1212}, 044 (2012),
  \href{http://arxiv.org/abs/arXiv:1210.2590}{{\tt arXiv:1210.2590}}.
  O.~Aharony, L.~Berdichevsky, M.~Berkooz and I.~Shamir,
  {\it Near-horizon solutions for D3-branes ending on 5-branes},
  Phys.\ Rev.\ D {\bf 84}, 126003 (2011),
  \href{http://arxiv.org/abs/arXiv:1106.1870}{{\tt arXiv:1106.1870}}.
  E.~D'Hoker, J.~Estes, M.~Gutperle and D.~Krym,
  {\it Exact Half-BPS Flux Solutions in M-theory. I: Local Solutions},
  JHEP {\bf 0808}, 028 (2008),
  \href{http://arxiv.org/abs/arXiv:0806.0605}{{\tt arXiv:0806.0605}}.

\bibitem{Gaiotto:2008ak} 
  D.~Gaiotto and E.~Witten,
  {\it S-Duality of Boundary Conditions In N=4 Super Yang-Mills Theory},
  Adv.\ Theor.\ Math.\ Phys.\  {\bf 13}, no. 3, 721 (2009),
  \href{http://arxiv.org/abs/arXiv:0807.3720}{{\tt arXiv:0807.3720}}.

\bibitem{Leblond:2001gn}
  F.~Leblond, R.~C.~Myers and D.~C.~Page,
  {\it Superstars and giant gravitons in M theory},
  JHEP {\bf 0201} (2002) 026,
  \href{http://arxiv.org/abs/hep-th/0111178}{{\tt hep-th/0111178}}.






 
\bibitem{Klebanov:1998hh} 
  I.~R.~Klebanov and E.~Witten,
  {\it Superconformal field theory on three-branes at a Calabi-Yau singularity},
  Nucl.\ Phys.\ B {\bf 536}, 199 (1998),
  \href{http://arxiv.org/abs/hep-th/9807080}{{\tt hep-th/9807080}}.
  

  
  
\bibitem{Bah:2013aha}
  I.~Bah and N.~Bobev,
  {\it Linear quivers and $ \mathcal{N} $ = 1 SCFTs from M5-branes},
  JHEP {\bf 1408} (2014) 121,
  \href{http://arxiv.org/abs/arXiv:1307.7104}{{\tt arXiv:1307.7104}}.
  
   

  \bibitem{Witten:1997sc}
  E.~Witten,
  {\it Solutions of four-dimensional field theories via M theory},
  Nucl.\ Phys.\ B {\bf 500} (1997) 3,
  \href{http://arxiv.org/abs/hep-th/9703166}{{\tt hep-th/9703166}}.
\bibitem{Gaiotto:2009we}
  D.~Gaiotto,
  {\it N=2 dualities},
  JHEP {\bf 1208} (2012) 034,
  \href{http://arxiv.org/abs/arXiv:0904.2715}{{\tt arXiv:0904.2715}}.
  

\bibitem{Bah:2011vv}
  I.~Bah, C.~Beem, N.~Bobev and B.~Wecht,
  {\it AdS/CFT Dual Pairs from M5-Branes on Riemann Surfaces},
  Phys.\ Rev.\ D {\bf 85} (2012) 121901,
  \href{http://arxiv.org/abs/arXiv:1112.5487}{{\tt arXiv:1112.5487}}.
  
\bibitem{Bah:2012dg}
  I.~Bah, C.~Beem, N.~Bobev and B.~Wecht,
  {\it Four-Dimensional SCFTs from M5-Branes},
  JHEP {\bf 1206} (2012) 005,
  \href{http://arxiv.org/abs/arXiv:1203.0303}{{\tt arXiv:1203.0303}}.
 
 
 

  
  
  
  
  
\bibitem{Benini:2009mz} 
  F.~Benini, Y.~Tachikawa and B.~Wecht,
  {\it Sicilian gauge theories and N=1 dualities},
  JHEP {\bf 1001}, 088 (2010),
  \href{http://arxiv.org/abs/arXiv:0909.1327}{{\tt arXiv:0909.1327}}.
  
  
\bibitem{Bah:2011je}
  I.~Bah and B.~Wecht,
  {\it New N=1 Superconformal Field Theories In Four Dimensions},
  JHEP {\bf 1307} (2013) 107,
  \href{http://arxiv.org/abs/arXiv:1111.3402}{{\tt arXiv:1111.3402}}.





\bibitem{Hanany:1996ie} 
  A.~Hanany and E.~Witten,
  {\it Type IIB superstrings, BPS monopoles, and three-dimensional gauge dynamics},
  Nucl.\ Phys.\ B {\bf 492}, 152 (1997),
  \href{http://arxiv.org/abs/hep-th/9611230}{{\tt hep-th/9611230}}.


\bibitem{Dasgupta:1998su}
  K.~Dasgupta and S.~Mukhi,
  {\it Brane constructions, conifolds and M theory},
  Nucl.\ Phys.\ B {\bf 551} (1999) 204,
  \href{http://arxiv.org/abs/hep-th/9811139}{{\tt hep-th/9811139}}.


\bibitem{Uranga:1998vf}
  A.~M.~Uranga,
  {\it Brane configurations for branes at conifolds},
  JHEP {\bf 9901} (1999) 022,
  \href{http://arxiv.org/abs/hep-th/9811004}{{\tt hep-th/9811004}}.
  
  
\bibitem{Gauntlett:2004zh}
  J.~P.~Gauntlett, D.~Martelli, J.~Sparks and D.~Waldram,
  {\it Supersymmetric AdS(5) solutions of M theory},
  Class.\ Quant.\ Grav.\  {\bf 21} (2004) 4335,
  \href{http://arxiv.org/abs/hep-th/0402153}{{\tt hep-th/0402153}}.    
  
  
\bibitem{Klebanov:2007ws}
  I.~R.~Klebanov, D.~Kutasov and A.~Murugan,
  {\it Entanglement as a probe of confinement},
  Nucl.\ Phys.\ B {\bf 796} (2008) 274,
  \href{http://arxiv.org/abs/arXiv:0709.2140}{{\tt arXiv:0709.2140}}.

  
\bibitem{Gubser:1998vd}
  S.~S.~Gubser,
  {\it Einstein manifolds and conformal field theories},
  Phys.\ Rev.\ D {\bf 59} (1999) 025006,
  \href{http://arxiv.org/abs/hep-th/9807164}{{\tt hep-th/9807164}}.
  

  
\bibitem{Myers:1999ps}
  R.~C.~Myers,
  {\it Dielectric branes},
  JHEP {\bf 9912} (1999) 022
  \href{https://arxiv.org/abs/hep-th/9910053}{{\tt hep-th/9910053}}.
  
  
  
  
\bibitem{Intriligator:2003jj}
  K.~A.~Intriligator and B.~Wecht,
  {\it The Exact superconformal R symmetry maximizes a},
  Nucl.\ Phys.\ B {\bf 667} (2003) 183,
  \href{http://arxiv.org/abs/hep-th/0304128}{{\tt hep-th/0304128}}.
  
  
  
\bibitem{Anselmi:1997am}
  D.~Anselmi, D.~Z.~Freedman, M.~T.~Grisaru and A.~A.~Johansen,
  {\it Nonperturbative formulas for central functions of supersymmetric gauge theories},
  Nucl.\ Phys.\ B {\bf 526} (1998) 543,
  \href{http://arxiv.org/abs/hep-th/9708042}{{\tt hep-th/9708042}}.
  

\bibitem{Tachikawa:2009tt}
  Y.~Tachikawa and B.~Wecht,
  {\it Explanation of the Central Charge Ratio 27/32 in Four-Dimensional Renormalization Group Flows between Superconformal Theories},
  Phys.\ Rev.\ Lett.\  {\bf 103} (2009) 061601,
  \href{http://arxiv.org/abs/arXiv:0906.0965}{{\tt arXiv:0906.0965}}.
  
  
\bibitem{Novikov:1983uc} 
  V.~A.~Novikov, M.~A.~Shifman, A.~I.~Vainshtein and V.~I.~Zakharov,
  {\it Exact Gell-Mann-Low Function of Supersymmetric Yang-Mills Theories from Instanton Calculus},
  Nucl.\ Phys.\ B {\bf 229}, 381 (1983).
  
\bibitem{ArkaniHamed:1997mj} 
  N.~Arkani-Hamed and H.~Murayama,
  {\it Holomorphy, rescaling anomalies and exact beta functions in supersymmetric gauge theories},
  JHEP {\bf 0006}, 030 (2000),
  \href{http://arxiv.org/abs/hep-th/9707133}{{\tt hep-th/9707133}}.
  

\bibitem{Cvetic:1999xp} 
  M.~Cvetic {\it et al.},
 {\it Embedding AdS black holes in ten-dimensions and eleven-dimensions},
  Nucl.\ Phys.\ B {\bf 558}, 96 (1999),
  \href{http://arxiv.org/abs/hep-th/9903214}{{\tt hep-th/9903214}}.



\bibitem{Apruzzi:2015zna}
  F.~Apruzzi, M.~Fazzi, A.~Passias and A.~Tomasiello,
  {\it Supersymmetric AdS$_{5}$ solutions of massive IIA supergravity},
  JHEP {\bf 1506} (2015) 195,
  \href{http://arxiv.org/abs/arXiv:1502.06620}{{\tt arXiv:1502.06620}}.

\bibitem{Bah:2015fwa}
  I.~Bah,
  {\it AdS5 solutions from M5-branes on Riemann surface and D6-branes sources},
  JHEP {\bf 1509} (2015) 163,
  \href{http://arxiv.org/abs/arXiv:1501.06072}{{\tt arXiv:1501.06072}}.


\bibitem{Duff:1997qz}
  M.~J.~Duff, H.~Lu and C.~N.~Pope,
  {\it Supersymmetry without supersymmetry},
  Phys.\ Lett.\ B {\bf 409} (1997) 136,
  \href{http://arxiv.org/abs/hep-th/9704186}{{\tt hep-th/9704186}}.


\bibitem{Elitzur:1997fh}
  S.~Elitzur, A.~Giveon and D.~Kutasov,
  {\it Branes and N=1 duality in string theory},
  Phys.\ Lett.\ B {\bf 400} (1997) 269,
  \href{http://arxiv.org/abs/hep-th/9702014}{{\tt hep-th/9702014}}.
  J.~T.~Liu and R.~Minasian,
  {\it Black holes and membranes in AdS(7)},
  Phys.\ Lett.\ B {\bf 457}, 39 (1999),
  \href{http://arxiv.org/abs/hep-th/9903269}{{\tt hep-th/9903269}}.


 



\end{thebibliography}

\end{document}